\documentclass[fleqn,usenatbib,longauth]{aa}

\usepackage{txfonts}
\usepackage[T1]{fontenc}

\DeclareRobustCommand{\VAN}[3]{#2}
\let\VANthebibliography\thebibliography
\def\thebibliography{\DeclareRobustCommand{\VAN}[3]{##3}\VANthebibliography}

\usepackage{graphicx}	% Including figure files
\usepackage{amsmath}	% Advanced maths commands
\usepackage{multirow}
\usepackage{url}
\usepackage{lineno}
\usepackage{xcolor}
\usepackage{adjustbox}
\usepackage[normalem]{ulem}

\usepackage{booktabs}
\usepackage{threeparttable}

\usepackage{placeins}

\usepackage{hyperref}
\hypersetup{
    unicode=false,                  % non-Latin characters in Acrobat book
    pdftoolbar=true,                % show Acrobat toolbar?
    pdfmenubar=true,                % show Acrobat menu?
    pdffitwindow=true,              % window fit to page when opened
    pdfstartview={FitH},            % fits the width of the page t
    pdftitle={The Fourth S-PLUS Data Release},          % title
    pdfauthor={Herpich},            % author
    pdfsubject={Astronomy},         % subject of the document
    pdfcreator={dvipdf},            % creator of the document
    pdfproducer={dvipdf},           % producer of the document
    pdfkeywords={}, % list of keywords
    pdfnewwindow=true,              % links in new window
    colorlinks=true,                % false: boxed links; true: colored
    linkcolor=red,                  % color of internal links
    citecolor=blue,                 % color of links to bibliography
    filecolor=magenta,              % color of file links
    urlcolor=blue,                  % color of external links
    breaklinks=true,
    linktocpage
}

\newcommand{\petro}{\texttt{Petro}\xspace}
\newcommand{\auto}{\texttt{auto}\xspace}

\newcommand{\iso}{\texttt{iso}\xspace}
\newcommand{\PSF}{\texttt{PSF}\xspace}
\newcommand{\PStotal}{\texttt{PSTotal}\xspace}
\newcommand{\single}{\texttt{single}\xspace}
\newcommand{\dual}{\texttt{dual}\xspace}
\newcommand{\ID}{\texttt{ID}\xspace}
\newcommand{\SN}{\textit{S/N}\xspace}
\newcommand{\rf}{RF}
\newcommand{\flex}{FlexCoDE}
\newcommand{\bnn}{BMDN}

\newcommand{\orcid}[1]{\href{https://orcid.org/#1}{\includegraphics[width=10pt]{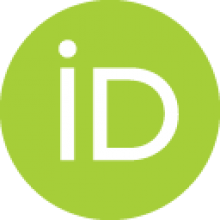}}}

\defcitealias{MendesDeOliveira+19}{MO19}
\defcitealias{AlmeidaFernandes+22}{AF22}

\begin{document}

%%%%%%%%%%%%%%%%%%%%%%%%%%%%%%%%%%%%%%%%%%%%%%%%%%

%%%%%%%%%%%%%%%%%%% TITLE PAGE %%%%%%%%%%%%%%%%%%%

% Title of the paper, and the short title which is used in the headers.
% Keep the title short and informative.
\title{The Fourth S-PLUS Data Release: 12-filter photometry covering $\sim$3000 square degrees in the southern hemisphere}

\titlerunning{The Fourth S-PLUS Data Release}
\authorrunning{The S-PLUS collaboration}

% The list of authors, and the shortlist which is used in the headers.
% If you need two or more lines of authors, add an extra line using \newauthor
\author{Herpich,~F. R.\inst{1}\fnmsep\thanks{email: fabio.herpich@ast.cam.ac.uk}\orcid{0000-0001-7907-7884},
Almeida-Fernandes,~F.\inst{2,3}\orcid{0000-0002-8048-8717},
Oliveira Schwarz,~G.~B.\inst{2,4}\orcid{0009-0003-6609-1582},
Lima,~E.~V.~R.\inst{2}\orcid{0000-0002-6268-8600},
Nakazono,~L.\inst{2,5}\orcid{0000-0001-6480-1155},
Alonso-García,~J.\inst{6,7}\orcid{0000-0003-3496-3772},
Fonseca-Faria,~M.~A.\inst{8}\orcid{0000-0002-7865-3971},
Sartori,~M.~J.\inst{2},
Bolutavicius,~G.~F.\inst{2}\orcid{0000-0001-7057-9685},
Fabiano de Souza,~G.\inst{2}\orcid{0000-0003-2426-8792},
Hartmann,~E.~A.\inst{9}\orcid{0000-0002-2527-8223},
Li,~L.\inst{2}\orcid{0000-0002-5295-7045},
Espinosa,~L.\inst{2},
Kanaan,~A.\inst{10}\orcid{0000-0002-2484-7551},
Schoenell,~W.\inst{11}\orcid{0000-0002-4064-7234},
Werle,~A.\inst{12}\orcid{0000-0002-4382-8081},
Machado-Pereira,~E.\inst{13}\orcid{0000-0001-6987-1531},
Gutiérrez-Soto,~L.~A.\inst{14}\orcid{0000-0002-9891-8017},
Santos-Silva,~T.\inst{2},
Smith Castelli,~A.~V.\inst{14,15},
Lacerda,~E.~A.~D.\inst{2}\orcid{0000-0001-7231-7953},
Barbosa,~C.~L.\inst{16}\orcid{0000-0002-4922-0552},
Perottoni,~H.~D.\inst{17}\orcid{0000-0002-0537-4146},
Ferreira Lopes,~C.~E.\inst{7,18},
Valença,~R.~R.\inst{2}\orcid{0000-0001-8847-0047},
Re Martho,~P.~A.\inst{2},
Bom,~C.~R.\inst{19,20},
Bonatto,~C.~J.\inst{21}\orcid{0000-0002-4102-1751},
Carvalho,~M.~S.\inst{2}\orcid{0009-0008-9042-4478},
Cernic,~V.\inst{2}\orcid{0000-0002-1233-0301},
Cid Fernandes,~R.\inst{10}\orcid{0000-0001-9672-0296},
Coelho,~P.\inst{2}\orcid{0000-0003-1846-4826},
Cortesi,~A.\inst{3},
Cubillos Palma,~B.\inst{22}\orcid{/0009-0007-0020-0976},
Doubrawa,~L.\inst{2}\orcid{0000-0001-8450-5193},
Ferreira Alberice,~V.~S.\inst{2,4},
Huaynasi,~F.~Q.\inst{13}\orcid{0000-0001-8741-8642},
Jacob Perin,~G.\inst{23}\orcid{0000-0002-5990-7497},
Jaque Arancibia,~M.\inst{22,24}\orcid{0000-0002-8086-5746},
Krabbe,~A.\inst{2},
Lima-Dias,~C.\inst{24}\orcid{0009-0006-0373-8168},
Lomelí-Núñez,~L.\inst{3}\orcid{0000-0003-2127-2841},
Lopes de Oliveira,~R.\inst{13,25}\orcid{0000-0002-6211-7226},
Lopes,~A.~R.\inst{14}\orcid{0000-0002-6164-5051},
Luiz Figueiredo,~A.\inst{2}\orcid{0009-0007-1625-8937},
Lösch,~E.\inst{2}\orcid{0000-0003-2561-0756},
Navarete,~F.\inst{26}\orcid{0000-0002-0284-0578},
Oliveira,~J.~M.\inst{2},
Overzier,~R.\inst{13,27,28}\orcid{0000-0002-8214-7617},
Placco,~V.~M.\inst{29}\orcid{0000-0003-4479-1265},
Roig,~F.~V.\inst{13}\orcid{0000-0001-7059-5116},
Rubet,~M.\inst{3}\orcid{0000-0001-9439-3121},
Santos,~A.\inst{19},
Sasse,~V.~H.\inst{10}\orcid{0009-0008-8763-7050},
Thainá-Batista, J.\inst{10}\orcid{0009-0008-2216-9575},
Torres-Flores,~S.\inst{22},
Beers,~T.~C.\inst{30},
Alvarez-Candal,~A.\inst{13},
Akras,~S.\inst{31}\orcid{0000-0003-1351-7204},
Panda,~S.\inst{8},
Limberg,~G.\inst{2},
Nilo Castellón,~J.~L.\inst{22},
Telles,~E.\inst{13}\orcid{0000-0002-8280-4445},
Lopes,~P.~A.~A.\inst{3}\orcid{0000-0003-2540-7424},
Pardo Montaguth,~G.~D.\inst{22}\orcid{0009-0003-1364-3590},
Beraldo e Silva, L.\inst{32}\orcid{0000-0002-0740-1507},
Humire,~P.~K.\inst{2}\orcid{0000-0003-3537-4849},
Borges Fernandes,~M.\inst{13}\orcid{0000-0001-5740-2914},
Cordeiro,~V.\inst{13}\orcid{0000-0001-9079-9511},
Ribeiro,~T.\inst{33}\orcid{0000-0002-0138-1365},
Mendes de Oliveira,~C.\inst{2}\orcid{0000-0002-5267-9065}\fnmsep\thanks{email: claudia.oliveira@iag.usp.br}
}

\institute{Cambridge Survey Astronomical Unit (CASU), Insitute of Astronomy, University of Cambridge, Madingley Road, Cambridge, CB3 0HA, GB\\\email{herpich@ast.cam.ac.uk}\and
Instituto de Astronomia, Geofísica e Ciências Atmosféricas, Rua do Matão, 1226, Cidade Universitária, São Paulo, 05508-090, BR\and
Observatório do Valongo, Ladeira Pedro Antônio, 43, Saúde, Rio de Janeiro, 20080-090, BR\and
Universidade Presbiteriana Mackenzie, Rua da Consolação, 930, Consolação, São Paulo, 01302-907, BR\and
University of Washington, 1415 NE 45th St, Seattle, 98195, US\and
Centro de Astronomía (CITEVA), Universidad de Antofagasta, Av. Angamos 601, Antofagasta, 1240000, CL\and
Millennium Institute of Astrophysics, Nuncio Monse\~nor Sotero Sanz 100, Of. 104, Providencia, 750-0000, CL\and
Laboratório Nacional de Astrofísica, Rua Estados Unidos, 154, Itajubá, 37504-364, BR\and
Instituto de Astrofísica de Canarias, C/ Vía Láctea s/n, La Laguna, 38205, ES\and
Universidade Federal de Santa Catarina, Campus Universitário Reitor João David Ferreira Lima, Florianópolis, 88040-900, BR\and
Giant Magellan Telescope, 251 S. Lake Ave, Suite 300, Pasadena, 91101, US\and
INAF - Osservatorio Astronomico di Padova, Vicolo dell'Osservatorio, 5, Padova, 35122, IT\and
Observatório Nacional, Rua General José Cristino, 77, Bairro São Cristóvão, Rio de Janeiro, 20921-400, BR\and
Instituto de Astrofísica de La Plata, UNLP-CONICET, Paseo del Bosque s/n, La Plata, B1900FWA, AR\and
Facultad de Ciencias Astronómicas y Geofísicas, Universidad Nacional de La Plata, Paseo del Bosque s/n, La Plata, B1900FWA, AR\and
Centro Universitário FEI, Av. Humberto de Alencar Castelo Branco, 3972, São Bernardo do Campo, 09850-901, BR\and
Nicolaus Copernicus Astronomical Center, Polish Academy of Sciences, Bartycka 18, Warsaw, 00-716, PL\and
Instituto de Astronomía y Ciencias Planetarias, Universidad de Atacama, Copayapu 485, Copiapó, 1530000, CL\and
Centro Brasileiro de Pesquisas Físicas, Rua Dr. Xavier Sigaud, 150, Rio de Janeiro, 22290-180, BR\and
Centro Federal de Educação Tecnológica Celso Suckow da Fonseca, Rodovia Mário Covas, lote J2, quadra J, Itaguaí, 23810-000, BR\and
Universidade Federal do Rio Grande do Sul, Av. Bento Gonçalves, 9500, Porto Alegre, 91501-970, BR\and
Departamento de Astronomía, Universidad de La Serena, Raúl Bitrán 1305, La Serena, , CL\and
Instituto de Matemática e Estatística, Rua do Matão, 1010, Cidade Universitária, São Paulo, 05508-090, BR\and
Instituto Multidisciplinario de Investigación y Postgrado, Universidad de La Serena, Raúl Bitrán 1305, La Serena, 1720236, CL\and
Departamento de Física, Universidade Federal de Sergipe, Av. Marechal Rondon, s/n, São Cristóvão, 49100-000, BR\and
Southern Astrophysical Research Telescope, Cerro Pachón, La Serena, 1720236, CL\and
Leiden Observatory, University of Leiden, Niels Bohrweg 2, Leiden, 2333 CA, NL\and
TNO, Oude Waalsdorperweg 63, Den Haag, 2597 AK, NL\and
NSF NOIRLab, 950 N. Cherry Ave, Tucson, 85719, US\and
Department of Physics and Astronomy and JINA Center for the Evolution of the Elements (JINA-CEE), 225 Nieuwland Science Hall, Notre Dame, 46556, US\and
Institute for Astronomy, Astrophysics, Space Applications and Remote Sensing, National Observatory of Athens, I. Metaxa \& Vas. Pavlou, Palaia Penteli, 15236, GR\and
Department of Astronomy \& Steward Observatory, University of Arizona, 933 N. Cherry Ave, Tucson, 85721, US\and
Rubin Observatory Project Office, 950 N. Cherry Ave, Tucson, 85719, US
}

% These dates will be filled out by the publisher
\date{Accepted XXX. Received YYY; in original form ZZZ}

% Enter the current year, for the copyright statements etc.
% \pubyear{2023}

% \label{firstpage}
% \pagerange{\pageref{firstpage}--\pageref{lastpage}}

\abstract
{The Southern Photometric Local Universe Survey (S-PLUS) is a project to map $\sim9300$ sq deg of the sky using twelve bands (seven narrow and five broadbands). Observations are performed with the T80-South telescope, a robotic telescope located at the Cerro Tololo Observatory in Chile. The survey footprint consists of several large contiguous areas, including fields at high and low galactic latitudes, and towards the Magellanic Clouds. S-PLUS uses fixed exposure times to reach point source depths of about $21$ mag in the $griz$ and $20$ mag in the $u$ and the narrow filters.}
{This paper describes the S-PLUS Data Release 4 (DR4), which includes calibrated images and derived catalogues for over 3000 sq deg, covering the aforementioned area. The catalogues provide multi-band photometry performed with the tools \texttt{DoPHOT} and \texttt{SExtractor} -- point spread function (\PSF) and aperture photometry, respectively. In addition to the characterization, we also present the scientific potential of the data.}
{We use statistical tools to present and compare the photometry obtained through different methods. Overall we find good agreement between the different methods, with a slight systematic offset of 0.05\,mag between our \PSF and aperture photometry. We show that the astrometry accuracy is equivalent to that obtained in previous S-PLUS data releases, even in very crowded fields where photometric extraction is challenging. The depths of main survey (MS) photometry for a minimum signal-to-noise ratio $S/N = 3$ reach from  $\sim19.5$ for the bluer bands to $\sim21.5$ mag on the red. The range of magnitudes over which accurate \PSF photometry is obtained is shallower, reaching $\sim19$ to $\sim20.5$ mag depending on the filter. Based on these photometric data, we provide star-galaxy-quasar classification and photometric redshift for millions of objects.}
{We demonstrate the versatility of the data by presenting the results of a project to identify members of four Abell galaxy clusters in the Local Universe. The S-PLUS DR4 data allow for a reliable assessment of cluster membership out to a large radius corresponding to $5\times r_{200}$. The S-PLUS DR4 can be accessed through the survey data portal. All the software used to generate the catalogues for this release and the scientific investigation presented is available in the collaboration GitHub repository.}
{The S-PLUS DR4 consists of a large, calibrated public dataset, providing powerful ways for studying Galactic and extra-galactic objects through an extensive set of (broad and narrow) filters.}

% \begin{keywords}
% keyword1 -- keyword2 -- keyword3
% \end{keywords}

\maketitle
%%%%%%%%%%%%%%%%%%%%%%%%%%%%%%%%%%%%%%%%%%%%%%%%%%

%%%%%%%%%%%%%%%%% BODY OF PAPER %%%%%%%%%%%%%%%%%%

%% Introduction %%
\section{Introduction}
\label{sec:intro}
%----------------------------------------------------------------
%1 Começar listando os objetivos do s-plus, citando o que já foi feito e o que foi alcançado no DR4

The Southern Photometric Local Universe Survey (S-PLUS) is a wide-area, multi-filter, ongoing photometric survey aiming to observe nearly 9300 sq deg in the sky using the T80-South robotic telescope located at the Cerro Tololo Inter-American Observatory, Chile (CTIO). The instrumentation, strategies, and goals of the survey are described in \citet[][hereafter MO19]{MendesDeOliveira+19}. S-PLUS is one of the many projects that follow the success of the Sloan Digital Sky Survey \citep[SDSS; ][]{York+2000}, which opened a new era of large photometric surveys, such as Two Micron All Sky Survey \citep[2MASS;][]{Skrutskie+2006}, the Galaxy Evolution Explorer \citep[GALEX;][]{Morrissey+2007}, the VLT Survey Telescope ATLAS \citep{Shanks+2015}, the Panoramic Survey Telescope and Rapid Response System \citep[PanSTARRS;][]{Chambers+2016}, the Dark Energy Survey \citep[DES;][]{Abbott+2018}, the SkyMapper Southern Sky Survey \citep{Wolf+2018}, the Legacy Survey of Space and Time \citep[LSST; ][]{Ivezic+2019}, among many others. The different instrument designs, strategies, and telescope locations make these surveys unique in specific ways, each presenting its characteristics regarding the covered area, photometric depth, wavelength range, and observation cadence, making them complementary.

The most outstanding characteristic of S-PLUS is the use of the Javalambre 12-filter photometric system, which includes four SDSS-like ($griz$) bands, the Javalambre $u$-band ($uJAVA$, which we call $u$ for simplicity throughout this work), and seven narrow bands ($J0378$: [O\,{\sc iii}], $J0395$: Ca HK, $J0410$: H$\delta$, $J0430$: G-band, $J0515$: Mg-triplet, $J0660$: H$\alpha$, $J0861$: Ca-triplet) designed to capture key stellar spectral features \citep{MarinFranch+12}. In terms of instruments, S-PLUS is a duplicate of the Javalambre Photometric Local Universe Survey \citep[J-PLUS;][]{Cenarro2019}, which is contained in the framework of the multi-filter surveys of the Observatorio Astrof\'isico de Javalambre and includes the Javalambre Physics of the Accelerating Universe Astrophysical Survey \citep[][]{2019AAS...23338301D}. While similar to J-PLUS regarding the main design and equipment, the tiling, and the exposure times, S-PLUS follows a different strategy during observations by deploying the software \texttt{chimera} to automate them as well as for calibration with the use of the detection image (see Section \ref{sec::dr4}) for the dual-mode photometry. Since there are no plans for an S-PLUS-like project for the southern hemisphere at the moment, the latter will remain the wide-area survey using the biggest number of photometric filters in the near future.

The first S-PLUS Data Release (DR1; \citetalias{MendesDeOliveira+19}) covered 336 sq deg in the SDSS STRIPE\,82 region. The second data release (DR2) expanded the coverage to
more than 1000 sq deg by including observations in the high-latitude ($b > 15^{\circ}$) regions \citep[][hereafter AF22]{AlmeidaFernandes+22}. The third data release (DR3) further added about 700 sq deg, including observations around the Hydra supercluster \citep{2000A&AS..141..123K,1985A&AS...61...93H,1987AJ.....93.1338D,1992ApJ...385..421S,1995ApJ...446..457S,2003MNRAS.339..652K,2006A&A...447..133P}. DR3 shares the same data reduction and photometric calibration strategy as DR2, only extending the area of the latter\footnote{For this reason DR3 has not been presented in a standalone paper. For authors using DR3 data for publication, we recommend citing either this work and/or \citetalias{AlmeidaFernandes+22}.}. Still, both DR2 and DR3 deploy significant improvements relative to DR1, for which we refer the reader to \citetalias{AlmeidaFernandes+22} for a discussion on the differences.

The approximately 1800 sq deg available in previous releases, in addition to about 1200 sq deg that will be presented here, have provided important results in different areas of astronomy, most notably low-metallicity stars \citep{2021ApJ...912L..32P,Placco+22, 2023arXiv231017024P, Almeida-Fernandes+23}, characterization of high-velocity stars \citep{2024MNRAS.527.6173Q}, halo stars in the neighbourhood \citep{2021ApJ...912..147W} and stellar clusters in the Small Magellanic Cloud \citep{Souza+24}, detection of globular clusters \citep{2022MNRAS.510.1383B}, identification of multiple stellar populations in globular clusters \citep{Hartmann+22}, spectral energy distribution (SED) fitting, stellar populations and emission lines \citep{Thaina-Batista+23}, galaxy morphologies \citep{2021MNRAS.507.1937B, Bom+23}, galaxy cluster member selection \citep{Olave-Rojas+23}, and the galaxy environment \citep[clusters and groups][]{2021MNRAS.500.1323L,Werner+23,2023MNRAS.524.5340M,Lima-Dias+23}. These findings complement and extend the scope of the previous works already cited in \citetalias{AlmeidaFernandes+22}, further enhancing the overall impact of our research.

In this paper, we describe and characterize the S-PLUS Fourth Data Release (S-PLUS DR4), as well as the changes implemented in the pipeline that make it possible to obtain and calibrate the photometry in regions of high interstellar extinction parametrized in terms of $E (B-V)$, and high stellar density. The main additions to the footprint are $\sim300$ sq deg (150 fields) covering the Small and Large Magellanic Clouds and the bridge between them (hereafter, the \textit{MC} region) and $\sim340$ sq deg (171 fields) in the region of the Galactic plane ($l \in [220^{\circ}, 275^{\circ}$], $b \in [-15^{\circ}, 5^{\circ}]$, hereafter the \textit{Disk} region). With $\sim400$ sq deg added to the main area of the survey (henceforth, the \textit{Main} region), the total area of DR4 is of $\sim3000$ sq deg, composed of 1629 fields, each with a field-of-view of $\approx 2$ sq deg in the sky. It contains the observations from previous DRs, with the addition of data from the MC, the Disk, and 409 new fields, along with improvements to the photometric calibration. The whole dataset was (re)processed using the same reduction pipeline version and configuration as for DR2 and DR3 (presented in \citetalias{AlmeidaFernandes+22}), with improvements implemented primarily in the calibration pipeline.

The main changes implemented in DR4 regarding images and catalogues are described in Section~\ref{sec::dr4}, while details of the photometric calibration are presented in Section~\ref{sec:calibration}. In Section~\ref{sec:data_charac}, the characterization of the data is discussed. The content of the S-PLUS catalogues, as well as the Value Added Catalogues (VACs), are presented in Section~\ref{sec:catalogues}, while data access is described in Section~\ref{sec:data_access}. Finally, Section~\ref{sec:science_cases} showcases a science case made possible with the DR4 data. In Section~\ref{sec:summary} we summarize our findings.

%% The S-PLUS DR4 %%
\section{The S-PLUS DR4}
\label{sec::dr4}
%----------------------------------------------------------------

In this section, we present details of the images and catalogues of DR4 in the 12 optical bands: $u$, $J0378$, $J0395$, $J0410$, $J0430$, $g$, $J0515$, $r$, $J0660$, $i$, $J0861$, and $z$ (see Appendix \ref{ap:splusfilters} for description of the S-PLUS filter transmission curves along with the several parameters estimated for them).
The data set has been calibrated using a modified version of the calibration technique described in \citetalias{AlmeidaFernandes+22}. Below we list the features included to tackle the challenges faced upon the inclusion of crowded areas such as the MC region and disk fields:
\begin{itemize}
    \item \PSF photometry: Obtaining the aperture photometry of the most crowded regions is often not feasible, thus demanding techniques that better deal with source detection and flux measuring in these regions \citep{2018A&A...619A...4A}. We perform \PSF photometry on crowded S-PLUS images using the code \texttt{DoPHOT} (\citealt*{Schechter+1993}; \citealt{2012AJ....143...70A}), as detailed in Section~\ref{sec:psfphot}.
    \item Interstellar medium (ISM) extinction correction: The calibration pipeline now considers ISM extinction maps available in the literature \citep[][]{Schlegel+98,Gorski+20,GaiaDR2}. However, the final magnitudes are not corrected for ISM extinction. A column is provided with the dataset containing the reference value for the sources that the user can apply when needed (see Appendix \ref{apd:columns} for the columns description).
    \item Single mode photometry: Detections and measurements are obtained individually for each filter using \texttt{Source Extractor} \citep[][\texttt{SExtractor} hereafter]{Bertin+Arnouts1996}. The single-mode photometry is included in DR4 in addition to the dual-mode used in previous data releases, which is obtained using a reference image to obtain the detection list for which the fluxes will be extracted through all filters. The dual-mode photometry is still available as usual;
    \item Use of weight maps during aperture photometry: The photometry obtained using \texttt{SExtractor} now considers the individual uncertainties in each pixel, resulting in better-estimated magnitude errors.
    \item Improved detection image: The detection image is now produced using \texttt{SWARP} \citep{2002ASPC..281..228B}, and weight maps are considered for the \dual mode photometry. This image is the result of the combination of the $griz$ filters, as in former data releases.
    \item Data catalogues and columns: With the addition of single-mode aperture and \PSF photometry, the data structure was slightly modified for the new data release (see Section \ref{sec:catalogues}). This means the queries to download the data must be adjusted accordingly.
\end{itemize}

%----------------------------------------------------------------
\subsection{Observations and coverage}
%----------------------------------------------------------------

\begin{figure}
\begin{center}
\includegraphics[width=0.47\textwidth]{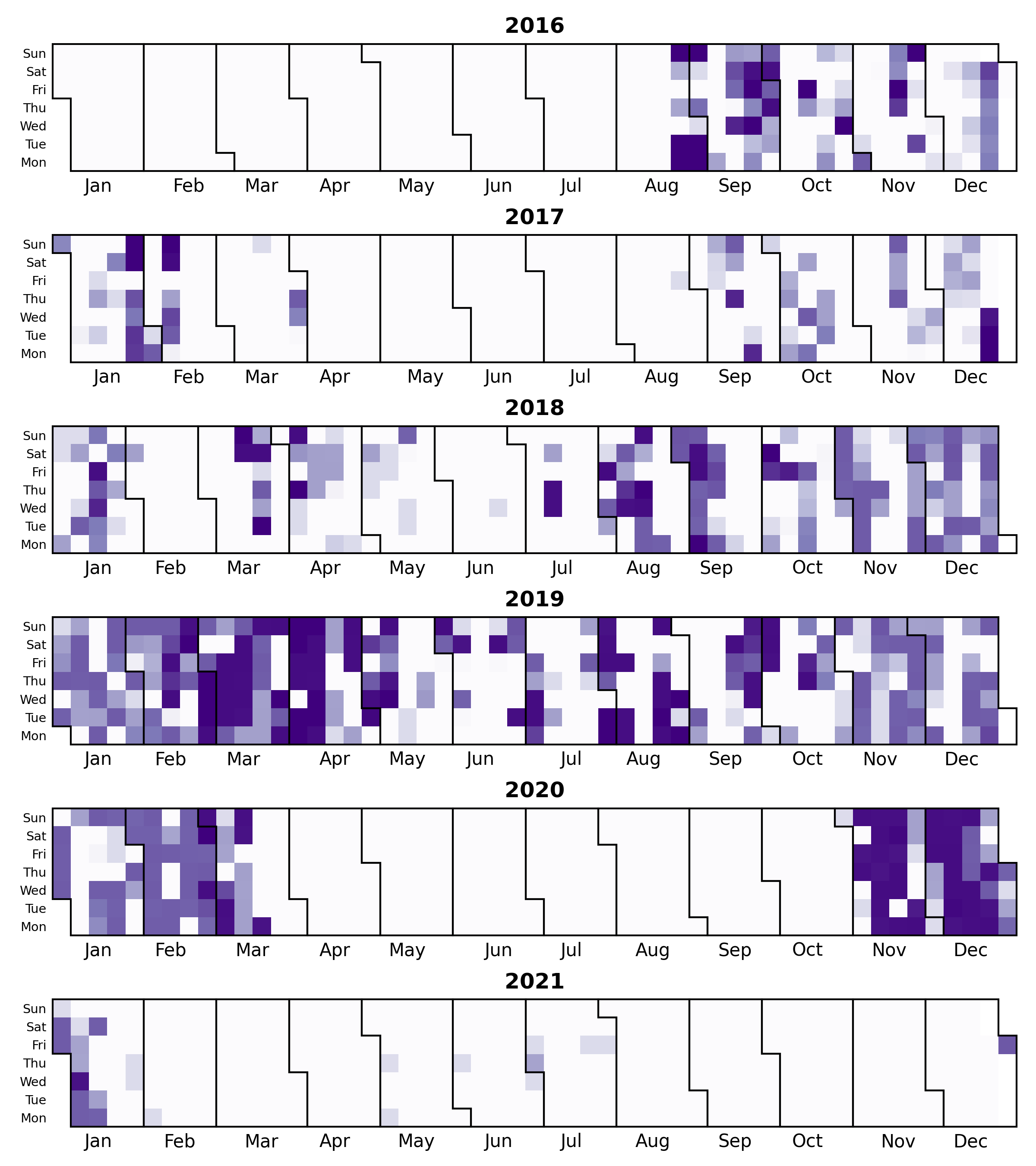}
\caption{Calendar of observations of the DR4 data. Blue squares represent nights where at least one observation occurred, with darker colours representing more data collected for a given night.}
\label{fig:obs_calendar}
\end{center}
\end{figure}

%The total area of DR4, 3022.7 sq deg, is shown as blue squares by Figure \ref{fig:footprint_iDR4}. 
The observations span a broad time window, the majority from August 2016 to February 2021 (with a few fields observed later in 2021). Fig.~\ref{fig:obs_calendar} shows the distribution of observations throughout the years. There were two significant interruptions: i) between April and November 2017, when technical issues were identified and subsequently fixed during the science verification phase, and ii) between March and October 2020, when COVID-19 restrictions forced the interruption of the observing activities. The DR4 contains observations of 1629 pointings (hereafter, fields) taken during 642 nights, corresponding to a total exposure time of $\sim2525$ hours (considering science images for the 12 filters). Fig.~\ref{fig:obs_calendar} shows nights where at least one DR4 observation occurred, with darker colours representing more observations during a particular night.

%----------------------------------------------------------------
%\begin{figure}
%\begin{center}
%\includegraphics[width=.48\textwidth]{figures/footprint_iDR4.png}
%\caption{The S-PLUS footprint. The blue filled squares represent the 1629 fields in the footprint of DR4. For reference, the \citet{Schlegel+98}}
%\label{fig:footprint_iDR4}
%\end{center}
%\end{figure}
%----------------------------------------------------------------
%----------------------------------------------------------------
%----------------------------------------------------------------
%The operation of the T80-South telescope is automatically performed by the observatory control system. Every night observations are planned to take into account the date, the moon's position, luminosity, and a set of predefined fields and sub-surveys priorities. During the observations, an automatic supervisor algorithm takes the weather conditions into account and executes the necessary changes to the observation schedule. Human interventions are also occasionally applied to ensure the best use of the telescope time.

\begin{figure}
\begin{center}
\includegraphics[width=0.47\textwidth]{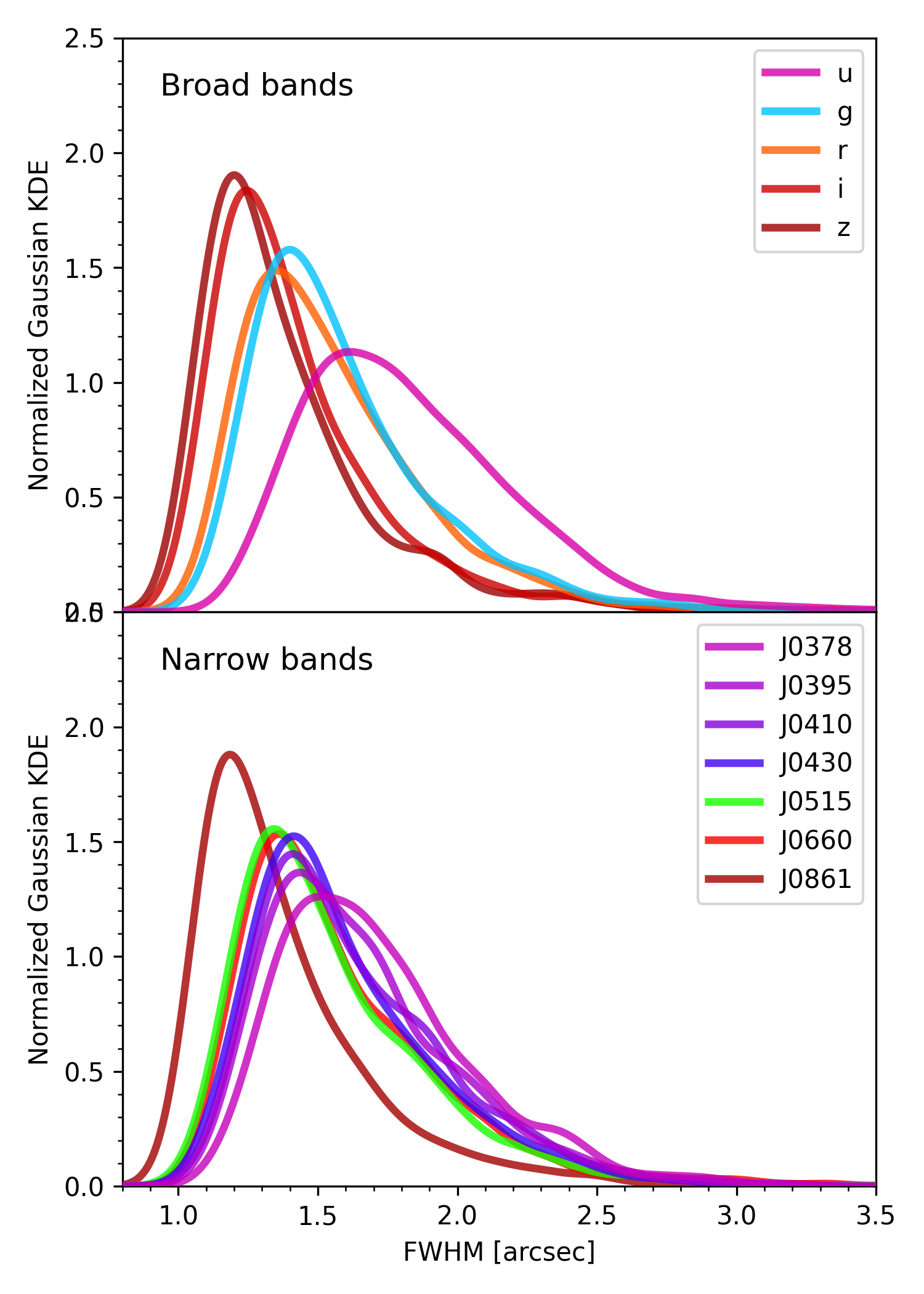}
\caption{Distribution of the characterized FWHM for the broadbands (top panel) and the narrow bands (bottom panel) for the whole DR4.}
\label{fig:iDR4_seeing}
\end{center}
\end{figure}

The S-PLUS MS is performed under good seeing conditions (typically smaller than 2\arcsec), dark or grey moon (moon $\mathrm{brightness} < 80$\%), and photometric conditions. This ensures the MS data set has the best quality that the T80-South can provide. Fig.~\ref{fig:iDR4_seeing} shows the distribution of mean full width at half maximum (FWHM), in arcsec, for the whole DR4. It is noticeable that bluer bands tend to have bigger FWHM. This effect is mainly due to the larger observations' exposure times, although other factors such as focus or distortions can play a role in this regard, especially on the borders.
%----------------------------------------------------------------

%----------------------------------------------------------------

%----------------------------------------------------------------
\subsection{Reduction and astrometry}

No changes relative to DR2 and DR3 were implemented in the reduction and astrometric calibration, which have already been described in \citetalias{AlmeidaFernandes+22}. In summary, the reduction of individual images follows the usual sequence of overscan subtraction, trimming, bias subtraction, masterflat division, cosmetic corrections (removing satellite tracks and cosmic rays), and fringing subtraction, as described in \citetalias{MendesDeOliveira+19} and \citetalias{AlmeidaFernandes+22}. In general, three dithered observations are coadded using \texttt{SWARP} to produce a single science image for each filter, although some tiles and/or filters my contain less or more than three observations, depending on technical and/or weather conditions at the time of acquisition. During the coadding, \texttt{SWARP} resamples the images and generates weight maps, providing the variance for each pixel. \texttt{SExtractor} uses these weight maps to estimate the uncertainties relative to the sources during the photometric estimation. The weight maps are not considered in the case of the \PSF photometry (see Section \ref{sec:psfphot}).

\begin{figure}
\begin{center}
\includegraphics[width=0.48\textwidth]{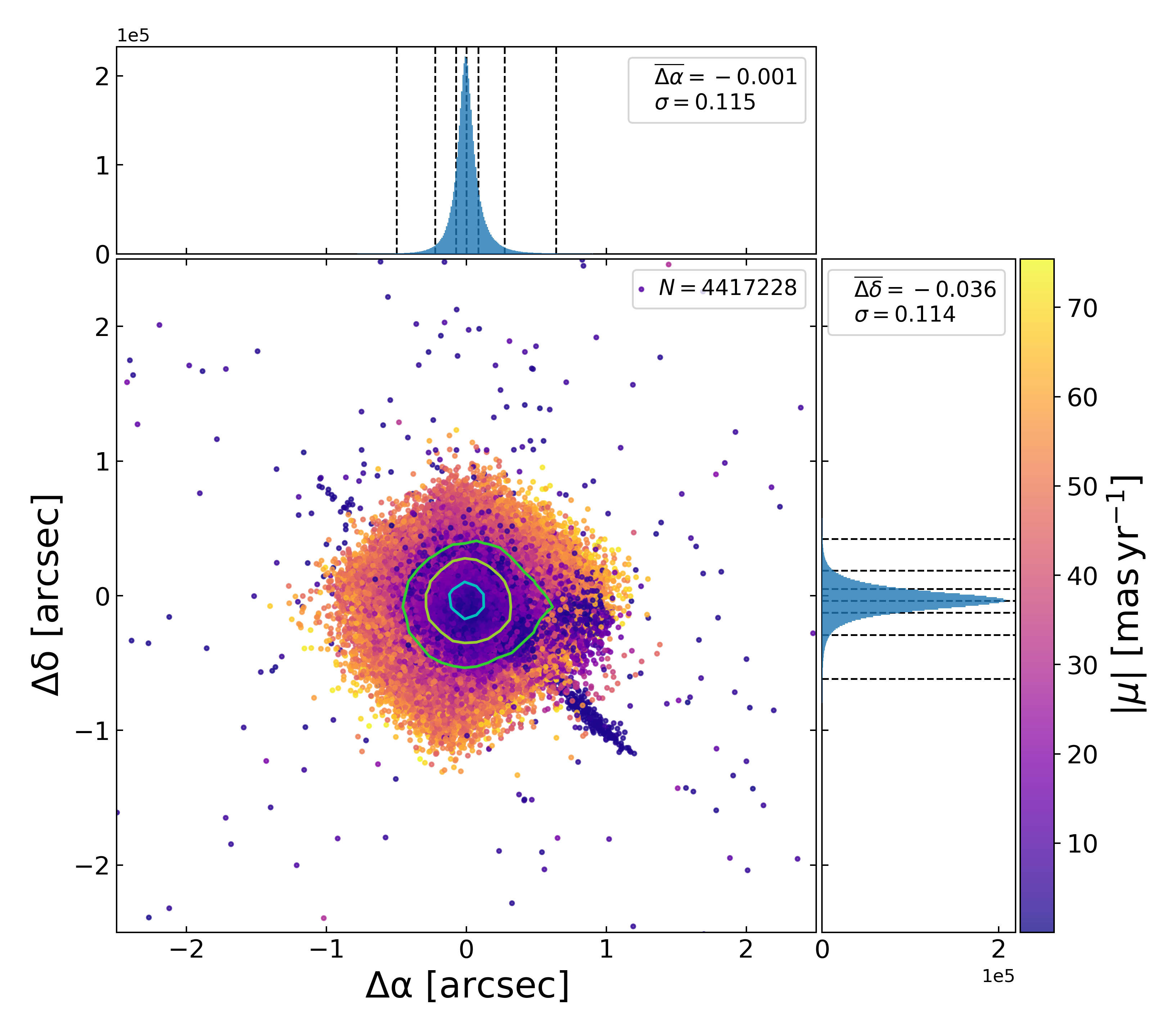}
\caption{Astrometric precision of the DR4 \PSF data compared to Gaia coordinates of over 4 million sources. Main panel: distribution of the sources colour-codded by the absolute value of the proper motion $|\mu|$ obtained from the Gaia data. The contours represent $1\sigma$ (cyan), $2\sigma$ (yellow), and $3\sigma$ (green) of the distribution. The histograms show the respective distributions for each axis with the dashed lines corresponding to $1\sigma$, $2\sigma$, and $3\sigma$. The histogram labels show the respective distribution's median and $\sigma$.}
\label{fig:Astrometry}
\end{center}
\end{figure}

The accuracy of the astrometric solution for high latitudes is the same as in \citetalias{AlmeidaFernandes+22}. For crowded fields (i.e. fields with over 300\,000 sources or smaller regions densely populated by stars and galaxies, like the centre of globular clusters, the inner parts of the Small and Large Magellanic Clouds and the fields of low Galactic latitude), however, the astrometry might get slightly worse due to difficulties in matching the sources with the reference catalogue, hence not being able to properly address the distortions of the images primarily at the edges of the field. This, in turn, can introduce more significant positional errors for the \PSF photometry, which we addressed by calculating the astrometric differences relative to Gaia Data Release 3 \citep[Gaia DR3;][]{GaiaDR3} similarly to those for the DR2 and DR3. The results are shown in Fig.~\ref{fig:Astrometry}, where the sources of \PSF DR4 $r$-band catalogues are compared with correspondent stars retrieved from Gaia data. The match is done using a radius of 5\arcsec\ and subsequent constraints are applied to remove high-proper motion stars, sources with parallax error $< 0.2$\arcsec, plus $14 < r < 18$\,mag, and $S/N > 20$. As for the proper motion, we discarded the 5\%\ of the objects with the higher absolute proper motions $|\mu|$, where $|\mu| = |\Delta\mathrm{RA}| + |\Delta\mathrm{Dec}|$, to avoid statistical interference from outliers that could affect our estimation (and this is the reason for the diamond-like shape of the scatter in Fig.~\ref{fig:Astrometry}). We notice that there are low-proper motion objects outside the main distribution region which we identified as outliers with a poorly defined centroid when obtaining their photometry. In general, they are related to fainter sources that are not necessarily detected in the same position depending on the filter considered. They are spread through the entire sample and correspond to less than 0.01\%\ of the sources. In the same analysis, we also see in Fig.~\ref{fig:Astrometry} a sample of correlated points forming a bar-like structure extending from the main scatter. These objects account for less than 0.02\%\ of the sample and are exclusively located at the corner of the images that compose the Large Magellanic Cloud's centre, where crowding increases the astrometric solution's errors. They are most likely associated with poorly defined centroids during photometry measuring due to internal astrometric inconsistencies.

%\subsection{Photometry}

%\PSF photometry is available for all 1629 fields in DR4 (crowded and uncrowded fields), while aperture photometry is available for 1416 fields (only the uncrowded fields have aperture photometry, given that this quantity does not make sense when stellar images overlap significantly).

%----------------------------------------------------------------
\subsection{Photometry}
%----------------------------------------------------------------

This section details the techniques employed for photometry determination in DR4. The photometry is obtained in several different ways, all of them with their pros and cons. At the first level, the two different photometry modes available are the aperture and \PSF, as described in Sections \ref{sec:aperturephot} and \ref{sec:psfphot}, respectively.

%The first is based on the sum of the counts inside a given aperture around the centroid of the detected source, while the latter consists of fitting a point spread function (hence, \PSF) to the light distribution around the detection centroid.  \PSF photometry works better in dense regions where the contamination from neighbouring sources makes the use of aperture photometry unfeasible

%As a result of the specific choices of input parameters used to reduce DR4 data,  we recommend that, when available, the PStotal aperture (which is the magnitude for a 3-arcsec aperture corrected to an aperture that contains all the stellar light) should be used as the best representation of the total magnitude of a point source.

%----------------------------------------------------------------
\subsubsection{Aperture photometry}\label{sec:aperturephot}
%----------------------------------------------------------------

Aperture photometry corresponds to adding up the detected flux within a specified radius centred on an object (in DR4, images are already sky-subtracted during the coadding process). This measured flux is then used to calculate an instrumental magnitude with \texttt{SExtractor}. Aperture photometry performs better in non-crowded fields, which is the case for most S-PLUS observations.

\bigskip
\noindent
\textit{Single and Dual modes}: 
The aperture photometry is further divided into two categories: i) in single mode, detections and measurements are independently obtained for each filter; and ii) in dual mode, all detections are first characterized using a reference `detection image' and the same centroids and apertures found for the sources are used for the measurements in all filters. In previous data releases, only dual-mode aperture photometry has been provided.

Using a detection image has the advantage of increasing the detection \SN while ensuring that the measurements in all filters for a given source are equivalent in position and aperture size (for the adaptive aperture). In S-PLUS, the detection image is obtained by coadding the images of the $griz$ filters, which, for DR4, is done with \texttt{SWARP} considering the individual weight maps for each filter. A weight map is also produced for the detection image at the end of this process.

Despite improving the detection thresholds, this dual approach can result in biases, favouring redder sources due to our choice of detection image. It is also not the best approach to detect sources that can be bright in a particular narrow band while being faint in the broadbands (e.g. H\,{\sc ii} regions with emission lines detectable in the $J0660$ filter), in which cases the single-mode aperture photometry is expected to perform better. The drawback, in this case, is that an extended source will not necessarily be detected in the same position, and will likely have adaptive apertures of different sizes for each filter, as the shape of the source may depend on the observed wavelength (a single source in one filter may even be split into multiple sources on another filter), reason why single-mode photometry should always be used with this in mind.

\bigskip
\noindent
\textit{Fixed and Adaptive apertures}: For both the single- and dual-mode photometries, the measurements are performed in several different apertures for each detected source. The \texttt{aper\_3} and \texttt{aper\_6} magnitudes correspond to measurements in fixed circular apertures of 3- and 6-arcsec diameter around the source. At the same time, the \auto and \petro are adaptive elliptical apertures based on the Kron and the Petrosian radius \citep{1976ApJ...209L...1P,1980ApJS...43..305K,Bertin+Arnouts1996}, respectively. The latter modes depend on the angular size and shape of the source, being more suitable for extended objects. The \iso aperture measures the flux based on the sum of the counts, but only for pixels attributed to a given source whose values are above a detection threshold. This better preserves the shape of an extended source at the cost of using fewer pixels to measure the flux \citep{Bertin+Arnouts1996}.

\bigskip
\noindent 
\textit{The aperture correction and the \PStotal magnitude}: In the final catalogues, we include only the 3- and 6-arcsec diameter fixed-circular apertures, but during the photometry estimation, we measure the magnitude at 32 different circular apertures between 1 and 50\arcsec\ in diameter. These apertures are used to build a growth curve for each filter at each observation (see \citetalias{AlmeidaFernandes+22}), representing how much the magnitude changes, on average, when increasing the aperture. For the estimation of the growth curve, we only use sources classified as stars (\texttt{CLASS\_STAR}$\geq 0.9$), which are neither too faint ($\SN \geq 30$) nor too bright ($\SN \leq~1000$). From this curve, the aperture correction is obtained relative to the 3\arcsec\ measure, corresponding to the value that must be added to this restricted value to get the total magnitude of the source. Since the growth curve directly results from the \PSF, this correction is only reliable for point sources (hence the name \PStotal). The \PStotal magnitude best represents the total magnitude of a point source while maintaining the high \SN of the very restricted 3-arcsec aperture. Whenever the aperture photometry is available for a given field, this is the instrumental magnitude used during the calibration process.

%----------------------------------------------------------------
\subsubsection{\PSF photometry}\label{sec:psfphot}
%----------------------------------------------------------------

\begin{figure}
\begin{center}
\includegraphics[width=0.47\textwidth]{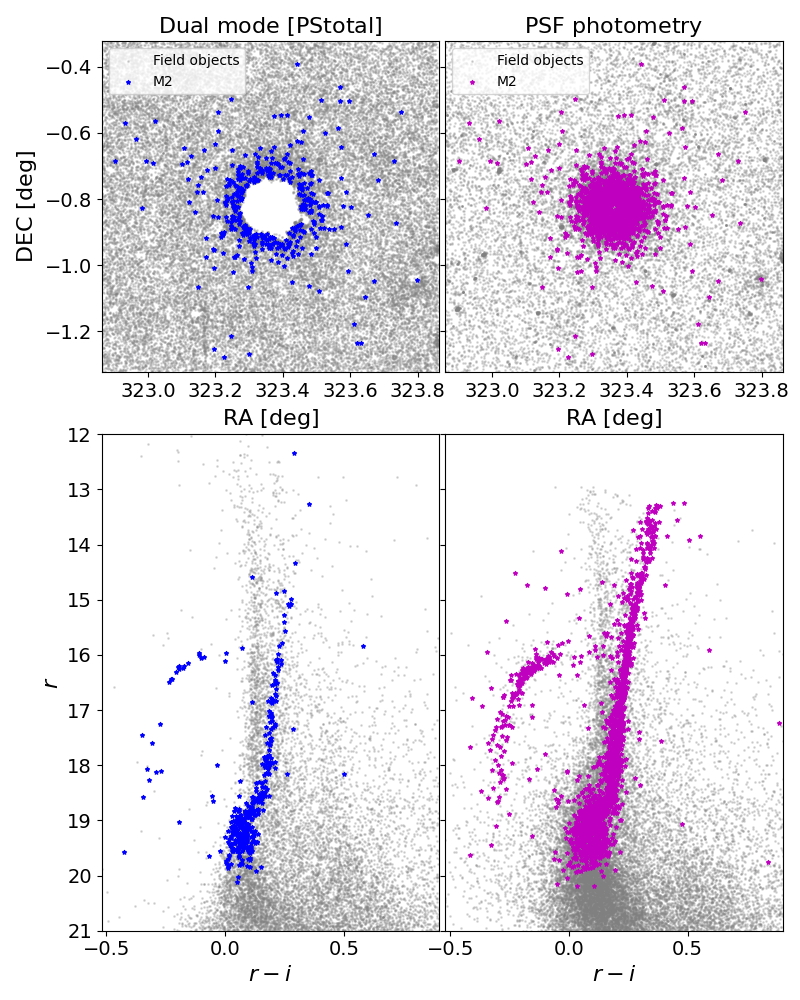}
\caption{Comparison between \PSF and aperture photometry in S-PLUS DR4. Top panels represent the spatial distribution of sources, while the bottom panels are colour-magnitude diagrams of the M2 globular cluster in the S-PLUS STRIPE82-0119 field. The blue points represent the sources detected using aperture photometry in the dual mode, and the magenta points are the ones obtained with the \PSF photometry. Background grey dots are field detections for the respective photometric mode.}
\label{fig:M2}
\end{center}
\end{figure}

\PSF photometry consists of fitting a point-spread function to the light distribution around the detection centroid. The DR4 \PSF photometry is measured using \texttt{DoPHOT} \citep{Schechter+1993,2012AJ....143...70A} and is available for all fields in DR4. The process of selecting the stars to be used as a reference to calculate the PSF of the observation is performed automatically by DoPHOT. Fig.~\ref{fig:M2}\footnote{The code used to select the sample and produce Fig.~\ref{fig:M2} is available on the S-PLUS official GitHub repository for the DR4 \url{https://github.com/splus-collab/codes-dr4-paper}.} shows the colour-magnitude diagram of the globular cluster Messier 2 (M2) located in the S-PLUS STRIPE82-0119\footnote{As listed in \citetalias{MendesDeOliveira+19}, S-PLUS main survey is divided in four named areas: Stripe82, Hydra cluster, MC, and ``Remaining S-PLUS''. By internal convention, the tiles/fields of each of these areas are named accordingly, e.g. STRIPE82-0$k$, HYDRA-0$k$, and MC0$k$, where $(k)_{001}^N$, with $N$ being the total number of tiles per area. The ``Remaining S-PLUS'' tiles are named SPLUS-$mjql$, where $m=n$ or $m=s$ if the tile is north or south of the galactic plane, respectively, $(j)_{01}^{L}$, where $L$ is the number of rows of tiles, $q=n$ or $q=s$ if the tile is north or south of the celestial equator, respectively, and $(l)_{01}^C$, where $C$ is the number of columns of tiles.} field. Fig.~\ref{fig:M2} was composed by selecting all stars within a region of 1 sq deg centred in the cluster coordinates for both \PStotal and \PSF modes. The selection of M2 members was based on the work of \citet{Hartmann+22} queried directly from Vizier\footnote{\url{https://vizier.cds.unistra.fr/viz-bin/VizieR-3?-source=J/MNRAS/515/4191/ngc7089}}, and matching the resulting catalogue with Gaia DR3 sources. With the proper motion measurements for the stars in the region, we limited the M2 sample to $\mu_\alpha = 3.51 \pm 0.5$\,mas/yr and $\mu_\delta = -2.16 \pm 0.5$\,mas/yr, where the mean values for the cluster were taken from \citet{2019MNRAS.482.5138B}. \PStotal photometry was further constrained to contain only objects with photometric flags 0, 1, 2 or 3\footnote{We refer the reader to the \texttt{SExtractor} manual \url{https://sextractor.readthedocs.io/en/latest/Flagging.html?highlight=flags\#extraction-flags-flags} for the description of the photometric flags.}. This comparison between the \PSF and the aperture (\PStotal) magnitudes shows how better is the PSF approach when working with denser regions.

%----------------------------------------------------------------
%----------------------------------------------------------------

Since the \PSF photometry is based on fitting the point-spread function to the light distribution of each source, it is only reliable as a measurement of the total flux for point sources, which can be selected according to the \texttt{CLASS\_STAR} parameter (see Section \ref{sec::class_star}). For the MC and Disk areas, where the aperture photometry is unreliable due to crowding, we adopt the instrumental \PSF magnitudes instead of \PStotal as the base instrumental magnitudes for the calibration. These correspond to 213 fields for which we only include \PSF photometry in DR4.

%----------------------------------------------------------------
\subsection{IDs}
%----------------------------------------------------------------

With the addition of single mode and \PSF photometry, the definition of a single ID for the same object in different filters becomes more complex and arbitrary. In \dual mode, the detection's centroid and adaptive aperture size are the same in all filters, so there is no ambiguity regarding what constitutes a source in each catalogue. For most applications, this \ID column is still sufficient to correctly identify each individual source. In the newly added modes, however, detections are independent in each filter, and, particularly for extended sources, a single object in one filter may be detected as multiple objects in another. Moreover, it may not be straightforward to identify the same source detected in different modes. For these reasons, an arbitrary definition of a source is necessary.

In DR4, three different levels of IDs are associated with the RA and Dec coordinates:
\begin{enumerate}
    \item \texttt{\{filter\}\_ID\_\{photometry\}}: Corresponds to the detection in a given filter and photometry that has the associated RA and Dec coordinates measured for the source. This ID is unique for the source only in a specific filter catalogue and has no relation to any other IDs in different filter/photometry catalogues. This ID cannot be used for any crossmatch. Example: ``DR4\_3\_STRIPE82-0001\_single\_r\_0000005'' is the \texttt{r\_ID\_single} for the 5th source detected by \texttt{SExtractor} in the single mode photometry for the field "STRIPE82-0001" in the $r$-band.
    
    \item \texttt{PHOT\_ID\_\{photometry\}}: For \dual mode IDs are obtained from the detection catalogue, which results from the detection image, and all sources are associated homogeneously across all filters. In the case of \single and \PSF photometry, this ID is generated by crossmatching the 12 filter catalogues with a tolerance of 3\arcsec, meaning that if a source has a position within this radius for multiple filters, the same \texttt{PHOT\_ID\_\{photometry\}} is assigned to it in all filters considered. This ID can be used for crossmatch between filters in the same photometry. Since this method combines measurements that can have different positional values, the corresponding \texttt{PHOT\_ID\_\{photometry\}\_RA} and \texttt{PHOT\_ID\_\{photometry\}\_DEC} is defined by the coordinates of the detection in the filter with the highest effective wavelength. Example: the same source of the previous example has a \texttt{PHOT\_ID\_SINGLE} of ``DR4\_3\_STRIPE82-0001\_single\_0114776'', which is the same for all 12 filters in \single mode photometry.
    
    \item \ID: This \ID is generated from the crossmatch between the \texttt{PHOT\_ID\_\{photometry\}} for all photometry modes available for a given field, with a tolerance of 5\arcsec. This means that a given source within five arcsec in the \PSF, \single and \dual mode photometries will be assigned to the same \ID. The corresponding \texttt{ID\_RA} and \texttt{ID\_DEC} are also arbitrarily defined as the coordinates of the \texttt{PHOT\_ID\_\{photometry\}} source, following a priority for detection in \dual > \single > \PSF. This \ID can be used for crossmatching between different photometries. Example: the source in the previous examples has an \ID of ``DR4\_3\_STRIPE82-0001\_0000001'' which, following the criteria described in this section, identifies this same source across the three photometry modes and all filters.
\end{enumerate}

For the three points above, \{filter\} is used as a replacement for any S-PLUS filter and \{photometry\} represents any of the photometric modes used, namely \dual, \single, and \PSF. While these different IDs are meant to facilitate the crossmatch between the different catalogues and allow access to different photometries for the same source, we highlight that the definition of what constitutes a single source is arbitrary, and our choices of the tolerance of 3\arcsec\ for the \texttt{PHOT\_ID\_\{photometry\}} and 5\arcsec\ for the \ID may not be ideal for all science cases. For the \single and \PSF photometries, in cases where these tolerances may be too large, we also recommend crossmatching between different filters/photometries using the \texttt{RA\_\{filter\}} and \texttt{DEC\_\{filter\}} coordinates listed in each catalogue.

% Photometric Calibration %%
\section{Photometric calibration}
\label{sec:calibration}
%----------------------------------------------------------------

The calibration of DR4 follows the same main steps as the previous data releases described in \citetalias{AlmeidaFernandes+22}, with some improvements highlighted in this section. S-PLUS magnitudes are calibrated to the AB system. At the same time, corrections are applied to ensure that the calibration of all fields is on the same scale, even when different calibration strategies are used in different regions. Since the template-fitting-based calibration is anchored in the magnitude of stars in the same observation as the science images, the air-mass correction is unnecessary once it is already included in the zero points (ZPs).

%----------------------------------------------------------------
\subsection{Calibration}
%----------------------------------------------------------------

The calibration method of the DR4 data is similar to what is described in detail in \citetalias{AlmeidaFernandes+22} for the DR2. However, for DR4, a subdivision in several more areas/strategies and additional steps were necessary.
The main steps can be summarized as follows:

\textit{External calibration}: A synthetic photometry library (see Section \ref{sec::syntlib}) is combined with a reference catalogue to predict the expected calibrated magnitudes for hundreds of selected sources in a field. ZPs are obtained from the difference between these predicted and the instrumental magnitudes.

\textit{Stellar locus calibration}: This step is only used for the Main$_D$, MC$_B$, and Disk$_A$ strategies (see Section \ref{sec:strategies} for an in-depth definition of the strategies), and is used to calibrate the filters $u$, $J0410$, and $J0430$ when the external calibration step is not available for one or more of these filters.

\textit{Internal calibration}: This step is a refinement of the previous calibration (from 0.01 to 0.02 mag), and takes advantage of the pre-calibrated narrow bands of S-PLUS to better constrain the synthetic models. For the strategies that use the stellar locus step (e.g. when calibration of the $u$ band is not feasible via the reference catalogue), filters $J0378$ and $J0395$ are only calibrated in this step.

\begin{figure}
\begin{center}
\includegraphics[width=0.47\textwidth]{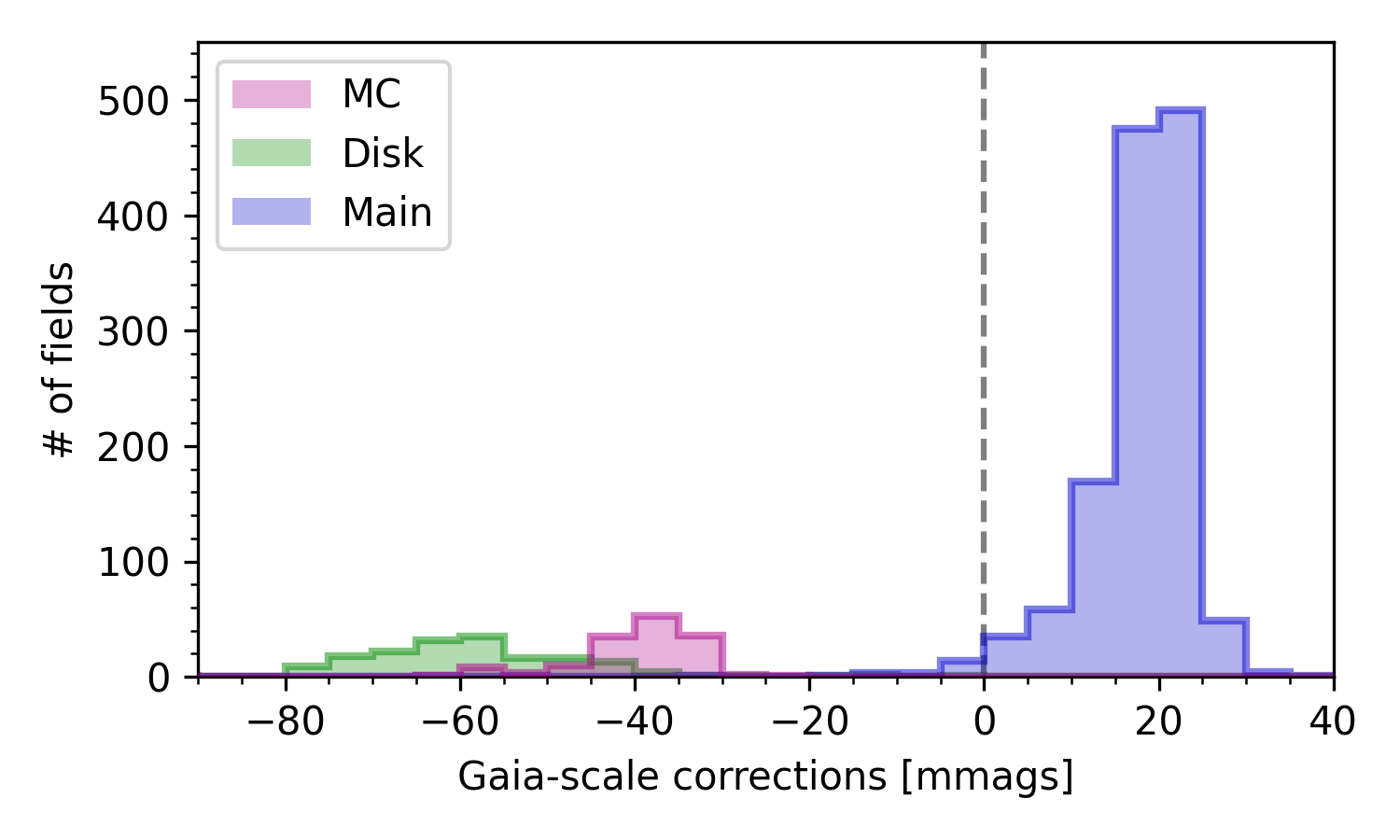}
\caption{Distribution of the Gaia-scale calibration ZPs for the fields in the main survey (blue), MC (pink), and disk (green).}
\label{fig:gaiascale_correction}
\end{center}
\end{figure}

\textit{Gaia-scale calibration}: In this final calibration stage, the synthetic library is combined with the S-PLUS magnitudes to predict the Gaia magnitudes ($BP$, $G$, $RP$) for hundreds of sources in the field. The average offset is applied to the ZP of each corresponding field for all data composing the release. Fig.~\ref{fig:gaiascale_correction} shows the distribution of the Gaia-scale offsets for different regions in DR4: the offset is of the order of $-60$, $-40$, and $20$ mmags for the fields in the Disk, MC, and MS, respectively.

%----------------------------------------------------------------
\subsection{Main improvements for DR4}
%----------------------------------------------------------------

The most important change is related to the ISM extinction. In DR2 and DR3, the grid of synthetic stellar photometry included models with ISM extinction, in terms of $E(B-V)$ \citep[][]{1994ApJ...429..582C}, varying between 0 and 1. However, some fields in DR4 can reach values as high as 5, and extending the grid while dealing with denser regions would result in prohibitive computational costs. Given this constraint, we chose to apply the extinction correction in the instrumental magnitudes used in the calibration. However, to account for errors in the extinction correction, we still allow some variation in the ISM extinction for the synthetic magnitudes. In the case of DR4, the grid range is $E(B-V) \in [-0.1, 0.1]$. The negative values allow the grid to still have suitable data points for cases when the extinction of the source is over corrected.

%----------------------------------------------------------------
\subsection{Synthetic photometry library}\label{sec::syntlib}
%----------------------------------------------------------------

The calibration pipeline relies on a synthetic photometry library to convert magnitudes between different filter sets by finding the model that best represents a source in a given system. This synthetic library was constructed by convolving the synthetic spectral library of \citet{Coelho14} with the transmission curves of all the different reference catalogues and the S-PLUS filter system. We adopted the extinction curve from \citet{Cardelli+89-v345-p245} with $R_V = 3.1$.

%----------------------------------------------------------------
\subsection{Calibration strategy}
\label{sec:strategies}
%----------------------------------------------------------------

The current version of the calibration relies on the chosen reference catalogue, the ISM extinction map, and the instrumental photometry used for the base of the calibration. We call each of these sets of configurations a calibration strategy. In the cases of DR2 and DR3, only three different strategies had to be applied. Still, the larger range of observational properties present in DR4 required the application of 10 different strategies, which are listed in Table~\ref{tab:calib_strategies}.

%----------------------------------------------------------------
\begin{table*}
\centering
    \begin{threeparttable}
    \caption{List of the different calibration strategies\tnote{$\dagger$} used in DR4.}
    \label{tab:calib_strategies}
      \begin{tabular}{@{}ccccc@{}}
        \toprule

        Calibration strategy & \# of fields & Reference & ISM Extinction map & \\ \midrule
        
        Main$_A$\tnote{a}  &  151  &  SDSS (Ivezic) [1]                                 &  Schlegel+98 [2] \\
        Main$_B$\tnote{a}  &  19   &  SDSS (DR12) [3]                                 &  Schlegel+98 [2] \\
        Main$_C$\tnote{a}  &  1050 &  ATLAS-REFCAT2 [4] + GALEX DR6/7 [5]   &  Schlegel+98 [2] \\
        Main$_D$\tnote{a}  &  82   &  ATLAS-REFCAT2 [4]                                 &  Schlegel+98 [2] \\
        Main$_E$\tnote{a}  &  4    &  Skymapper (DR1.1) [6]                            &  Schlegel+98 [2] \\
        MC$_A$\tnote{b}    &  111  &  ATLAS-REFCAT2 [4] + GALEX DR6/7 [5] (DR1.1) [6]  &  Schlegel+98 [2] \\
        MC$_B$\tnote{b}    &  20   &  ATLAS-REFCAT2 [4] (DR1.1) [6]                     &  Gorski+20 [7]   \\
        MC$_C$\tnote{b}    &  16   & Skymapper (DR1.1) [6]                            &  Schlegel+98 [2] \\
        MC$_D$\tnote{b}    &  3    & Skymapper (DR1.1) [6]                            &  Gorski+20 [7]   \\
        Disk$_A$\tnote{b}  &  174  &  ATLAS-REFCAT2 [4]                                 & Gaia (DR2) [8] \\
        \bottomrule

        \end{tabular}
      \begin{tablenotes}
        \small
       \item[\textdagger] References: [1] \citet{Ivezic+2007}, [2] \citet{Schlegel+98}, [3] \citet{Shadab+2015}, [4] \citet{Tonry+2018}, [5] \citet{Bianchi+2017}, [6] \citet{Wolf+2018}, [7] \citet{Gorski+20}, [8] \citet{GaiaDR2}.
        \item[a] Aperture (dual mode).
        \item[b] \PSF.
      \end{tablenotes}
    \end{threeparttable}
  \end{table*}
%----------------------------------------------------------------

\begin{figure*}
\begin{center}
\includegraphics[width=\textwidth]{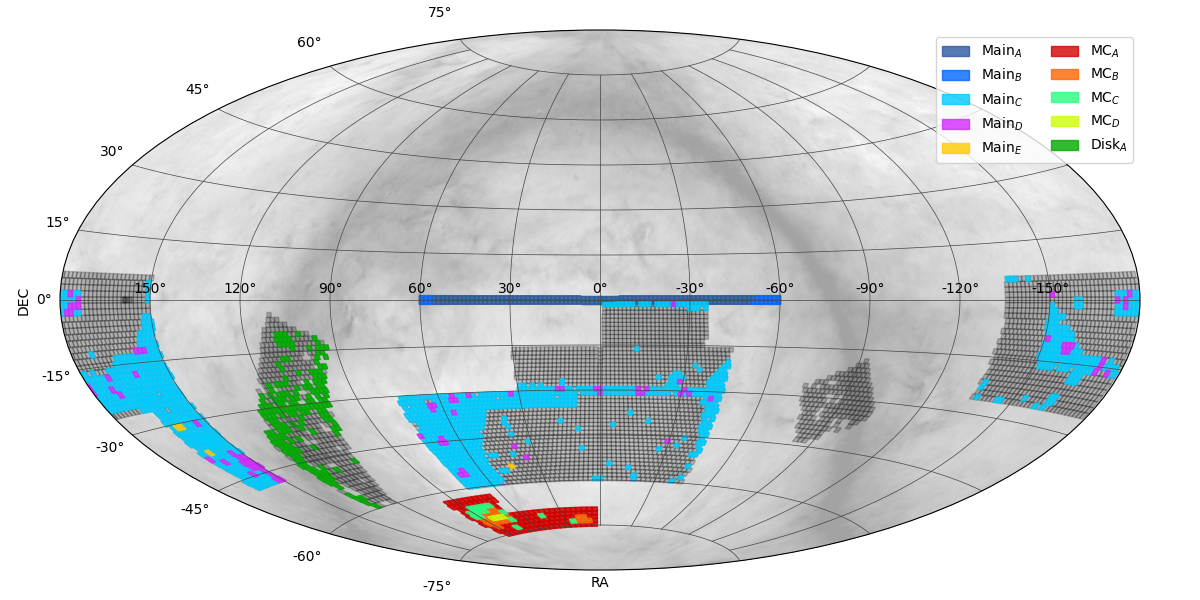}
\caption{Footprint of S-PLUS (grey squares) with the pointings of the DR4 coloured by the calibration strategy adopted for the calibration.}
\label{fig:footprint_strategies}
\end{center}
\end{figure*}

The choice of strategy depends on the availability of reference catalogues and extinction maps in the region and how crowded the fields are (e.g. low/high galactic fields). The base photometry corresponds to which instrumental magnitudes of S-PLUS were used as the base for the calibration. In most cases, the calibration is done relative to the \dual mode aperture photometry, but for crowded regions, the \PSF photometry is used instead. Fig.~\ref{fig:footprint_strategies} presents the spatial distribution for each calibration strategy.

%----------------------------------------------------------------
%----------------------------------------------------------------

\begin{figure}
\begin{center}
\includegraphics[width=0.47\textwidth]{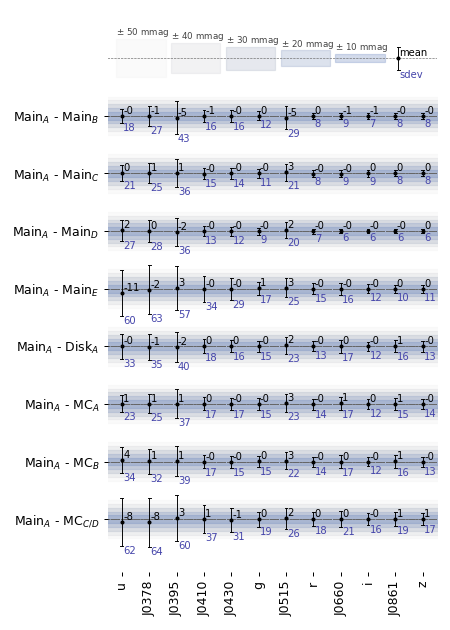}
\caption{Average differences between the ZP for the Main$_A$ and each strategy applied in 151 fields in the STRIPE82 region, for each of the 12 filters. The mean values for each distribution are shown in black near each data point. The error bar represents the standard deviation of the distribution, which is also listed in the blue text below each bar. Both sets of values are in mmag.} %For the MC$_D$ strategy, since the \citet{Gorski+20} extinction map is only available in the region of the Magellanic Clouds, the \citet{Schlegel+98} was used in the calibration comparison in the STRIPE82 region.
\label{fig:zp_comparison}
\end{center}
\end{figure}

The use of distinct strategies could result in offsets in the photometric calibration between different observations. To avoid this, we have calibrated the STRIPE82 region using all calibration strategies to estimate their average offset. The calibration of STRIPE82 using the \citet{Ivezic+2007} catalogue corresponds to the primary strategy used in the previous data releases and is considered the base calibration of S-PLUS. These offsets are corrected during the external calibration step. In Fig.~\ref{fig:zp_comparison}, we show the distribution of the differences in the ZPs for each strategy and filter compared to the Main$_A$ calibration for the DR4.

%----------------------------------------------------------------
%----------------------------------------------------------------

In general, the offsets (mean differences) are smaller than 10 mmags for all calibrations, and the scatter (standard deviation) is usually smaller than 10 mmags for the red filters ($g$ to $z$). For the blue filters ($u$ to $J0430$), the scatter is slightly bigger but still below 30 mmags in most cases. For the strategies Main$_E$, MC$_C$, and MC$_D$, this scatter in the blue bands can be as large as 60 mmags. However, these strategies are only applied to a few fields (4, 16, and 3, respectively), where the other strategies could not be applied.

% Data Characterization %%
\section{Data characterization}
\label{sec:data_charac}
%----------------------------------------------------------------

In this section, we present the characterization of the DR4. We discuss the ZP distribution, photometric depths, \SN, the completeness for each filter when compared to the $r$ band detections, and the AB calibration of the $u$ filter.

%----------------------------------------------------------------
\subsection{ZP distribution}
%----------------------------------------------------------------

\begin{figure}
\begin{center}
\includegraphics[width=0.47\textwidth]{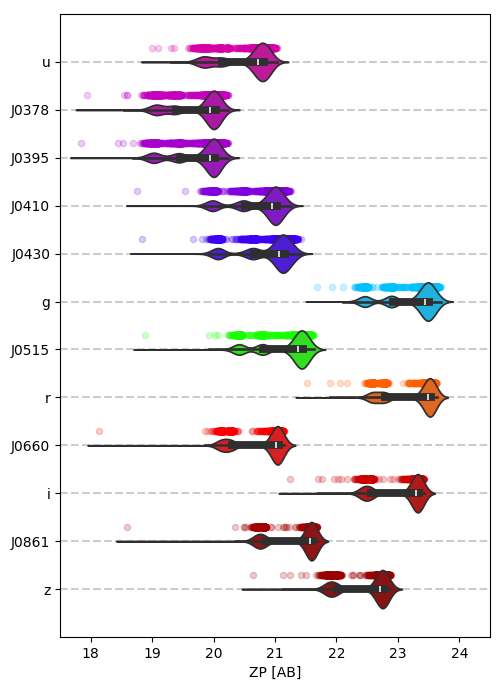}
\caption{Violin distribution of the ZP of all 1629 fields observed in the DR4 for each filter. The differences seen in any given filter distribution are mainly associated with the use of distinct detector configurations throughout the survey lifespan.}
\label{fig:zp_dist}
\end{center}
\end{figure}

Fig.~\ref{fig:zp_dist} shows the ZP distributions for all 1629 fields in DR4, where we can see from the violin-shaped distributions that the ZPs may differ as much as one magnitude for distinct fields/observations for the same filter. The main contributor to these variations is the use of different CCD configurations throughout the survey's history, which accounts for the doubled (and sometimes tripled) peak distributions in Fig.~\ref{fig:zp_dist}. Within each peak, the variation is caused by multiple factors, including the atmospheric conditions, observation airmass and variations in the instrument responses over time. All these effects are indirectly considered during the photometric calibration by ensuring that our magnitudes are calibrated to the same scale as a reference catalogue and later refined to the scale of the Gaia calibration, which is assumed to be uniform across the sky.

%----------------------------------------------------------------
%----------------------------------------------------------------

The Gaia-scale correction corresponds to the average difference between the real and the predicted Gaia magnitudes. The predicted magnitudes are determined by fitting the synthetic photometry to the 12 S-PLUS magnitudes (averaged for the three Gaia filters between hundreds of stars in each field). Even though, as mentioned above, the ZPs of all fields may vary by up to 1 mag (see Fig.~\ref{fig:zp_dist}), the corrections needed to bring the magnitude of all sources across the footprint to the Gaia sale are only approximately 20, 40 and 60 mmags for the MS, the MC, and the disk regions, respectively. This indicates that even with significant differences in the ZPs, the resulting magnitudes for the sources are within the expected precision of the observations.

The variations in the Gaia-scale correction as a function of region can be explained by the use of different base photometries and extinction maps between these calibrations. This correction is expected to bring all the calibrations to the same level.

%----------------------------------------------------------------
\subsection{Photometric depths}\label{sec:photdepth}
%----------------------------------------------------------------

We determined the photometric depths from the \petro magnitudes of objects in the MS's 1306 fields. This accurately represents the typical survey depth but does not include the Disk and MC, for which the depth is expected to be shallower (due to crowding). 

\begin{figure*}
\begin{center}
\includegraphics[width=1.96\columnwidth]{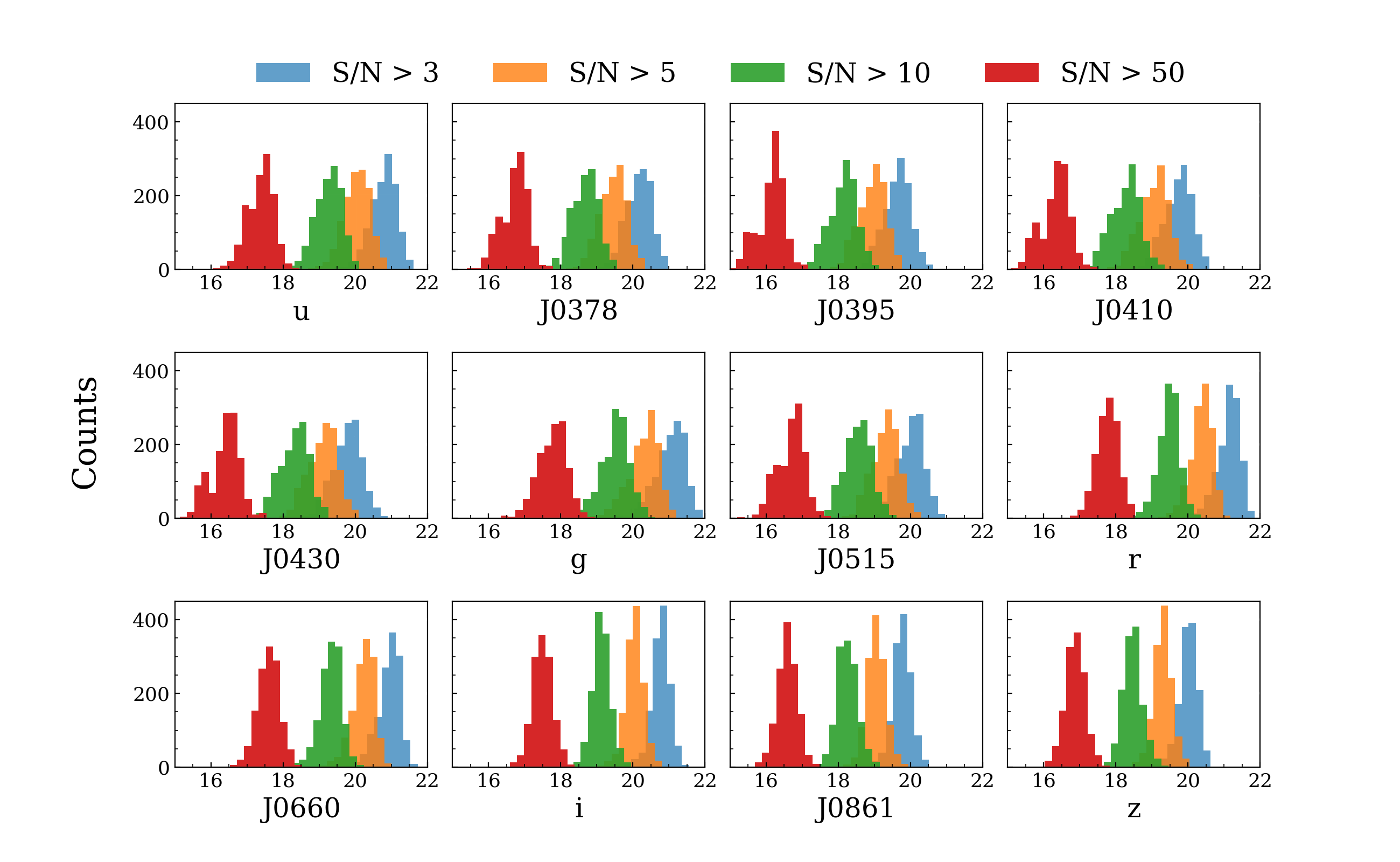}
\caption{12 panels showing the depths for one of the S-PLUS filters using the 1306 fields in the DR4 MS. Each histogram is constructed using the calculated photometric depths from the peak values of the \petro magnitude distribution for a given field, filter, and \SN threshold of 3, 5, 10, and 50 (blue, orange, green, and red, respectively).}
\label{fig:photdepths}
\end{center}
\end{figure*}

The measurements were done in the following way: for each field, the photometric depths were calculated for all 12 bands for \SN 3, 5, 10, and 50. Then, we accounted for the depths of all fields to characterize the depth of the MS. The kernel density estimator method is used to evaluate the peak of the magnitude distribution for each field, filter, and \SN threshold. This peak value represents the photometric depth, listed in Table~\ref{tab:photdepths} and shown by Fig.~\ref{fig:photdepths}. 

We observed that when the \SN is above 5, the depths for the blue filters ($u$, $J0378$, $J0395$) are approximately 20.1, 19.5, and 19 mag, reaching upper limits around 21, 20.4, and 19.7 mag, respectively. For $J0410$, $J0430$, and $g$, the respective depths are in the neighbourhood of 19.1, 19.2, and 20.4 mag, and the maximum values are around 20.1, 20.1, and 21.2 mag. As for $J0515$, $r$, and $J0660$, we obtained depths of 19.4, 20.4, and 20.3 mag, reaching upper limits around 20.3, 21.1, and 21 mag. Meanwhile, for the redder filters $i$, $J0861$, and $z$, we obtained approximately 20.1, 19.1, and 19.3 mag, with maximum values around 20.7, 20, and 20 mag, respectively.

%----------------------------------------------------------------
\begin{table}[ht]
\caption{Photometric depths from \petro magnitudes for different the \SN considered in a set of 1306 MS fields.}
\label{tab:photdepths}
\begin{tabular}{ccccc}
\toprule
Filter & $\text{SN} > 3$          & $\text{SN} > 5$          & $\text{SN} > 10$         & $\text{SN} > 50$  \\
\midrule
$u$      & $20.83_{-0.40}^{+0.30}$ & $20.07_{-0.39}^{+0.33}$ & $19.29_{-0.41}^{+0.33}$ & $17.41_{-0.43}^{+0.31}$ \\ [.4em]
$J0378$  & $20.19_{-0.39}^{+0.33}$ & $19.48_{-0.42}^{+0.34}$  & $18.69_{-0.44}^{+0.34}$ & $16.77_{-0.49}^{+0.29}$ \\ [.4em]
$J0395$  & $19.67_{-0.42}^{+0.32}$ & $18.99_{-0.44}^{+0.33}$  & $18.18_{-0.46}^{+0.33}$ & $16.21_{-0.54}^{+0.26}$ \\ [.4em]
$J0410$  & $19.78_{-0.45}^{+0.34}$ & $19.12_{-0.48}^{+0.33}$  & $18.31_{-0.49}^{+0.35}$ & $16.38_{-0.56}^{+0.31}$ \\ [.4em]
$J0430$  & $19.82_{-0.45}^{+0.35}$ & $19.15_{-0.50}^{+0.35}$  & $18.37_{-0.51}^{+0.34}$ & $16.46_{-0.59}^{+0.30}$ \\ [.4em]
$g$      & $21.10_{-0.47}^{+0.34}$ & $20.37_{-0.48}^{+0.30}$  & $19.54_{-0.44}^{+0.34}$ & $17.81_{-0.45}^{+0.34}$ \\ [.4em]
$J0515$  & $20.05_{-0.44}^{+0.31}$ & $19.35_{-0.44}^{+0.32}$  & $18.57_{-0.42}^{+0.34}$ & $16.74_{-0.50}^{+0.30}$ \\ [.4em]
$r$      & $21.18_{-0.36}^{+0.25}$ & $20.41_{-0.34}^{+0.24}$  & $19.50_{-0.32}^{+0.25}$ & $17.78_{-0.32}^{+0.30}$ \\ [.4em]
$J0660$  & $20.98_{-0.34}^{+0.25}$ & $20.26_{-0.34}^{+0.26}$  & $19.34_{-0.31}^{+0.25}$ & $17.61_{-0.33}^{+0.29}$ \\ [.4em]
$i$      & $20.79_{-0.24}^{+0.21}$ & $20.05_{-0.24}^{+0.22}$  & $19.13_{-0.24}^{+0.24}$ & $17.50_{-0.28}^{+0.26}$ \\ [.4em]
$J0861$  & $19.78_{-0.24}^{+0.24}$ & $19.05_{-0.25}^{+0.25}$  & $18.25_{-0.27}^{+0.29}$ & $16.59_{-0.28}^{+0.28}$ \\ [.4em]
$z$      & $20.03_{-0.25}^{+0.24}$ & $19.30_{-0.23}^{+0.24}$  & $18.46_{-0.25}^{+0.26}$ & $16.88_{-0.27}^{+0.27}$\\
\bottomrule
\end{tabular}%
% }
\end{table}
%----------------------------------------------------

% From the Main survey, We determined the photometric depths from the petro magnitudes of objects present in 1306 fields. This represents the typical survey depth well but does not include the Galactic Disk and Magellanic clouds, for which the depth is expected to be shallower (due to crowding). 

% The measurements were done the following way: for each field, the photometric depths were calculated for all 12 bands for four values of maximum S/N: less than 3, 5, 10 and 50. 

% The kernel density estimator method is used to evaluate the peak of the magnitude distribution for each field, filter and S/N threshold. This peak value represents the photometric depth and is listed in Table~\ref{tab:photdepths}, shown in Figure~\ref{fig:photdepths}. 

%----------------------------------------------------------------
%----------------------------------------------------------------

%----------------------------------------------------
% \subsection{Signal-to-noise}
\subsection{Completeness}
%----------------------------------------------------

\begin{figure*}
\begin{center}
\includegraphics[width=1.96\columnwidth]{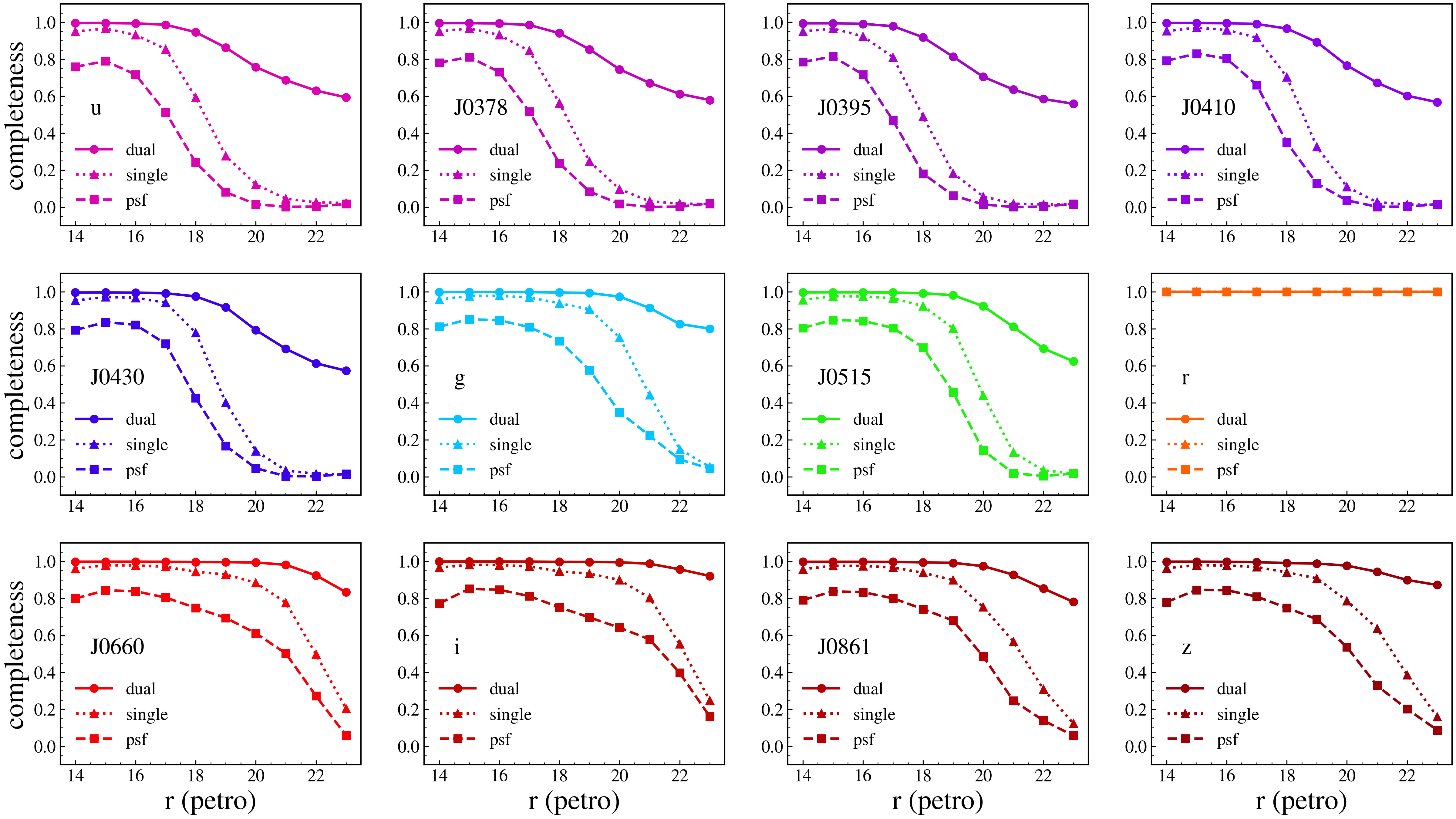}
\caption{Completeness of source detection in different filters relative to the $r$-band, the deeper of the twelve S-PLUS filters. Completeness is the fraction of sources detected in $r$ that are also seen in other filters ($ugiz$). Each panel corresponds to a filter and shows the completeness for \PSF, \single, and \dual photometry relative to the $r$ magnitude.}
\label{fig:completness}
\end{center}
\end{figure*}

Fig.~\ref{fig:completness} presents the completeness of photometric detections across various filters. The completeness is plotted against the magnitude in $r$ for \PSF, \single, and \dual photometry methods, revealing the impact of exposure times and filter sensitivities. Each filter's subplot demonstrates the proportion of sources detected in $r$ that are also observed in that filter.

%----------------------------------------------------------------
%----------------------------------------------------------------

For an additional representation of the total numbers associated with the detection, we refer the reader to the S-PLUS DR4 documentation\footnote{\url{https://splus.cloud/documentation/DR4}.}.

%----------------------------------------------------
 \subsection{AB calibration of the $u$-band}
%----------------------------------------------------

As an additional check for the accuracy of the calibration, we employed a Bayesian SED-fitting strategy to obtain estimates of atmospheric parameters ($T_{\mathrm{eff}}$, $\log{g}$, and [Fe/H]) for DR4 stars using as inputs all 12 S-PLUS bands. The method consists of maximizing the posterior considering a likelihood based on the difference between the input and the predicted magnitudes of the SEDs,
and a prior, based on the frequency of the parameters on a simulated sample of stars using MIST isochrones \citep{Dotter2016}. As an exercise, we used this process to check the accuracy of the broadband's calibration by replacing the magnitudes of one S-PLUS filter with the corresponding data from SDSS. The difference in the transmission curves was considered during the SED-fitting process. We repeated the process five times, replacing one of the broadbands in each iteration.

We noticed that replacing the $g$, $r$, $i$, and $z$ bands of S-PLUS with their counterparts from SDSS yields no significant difference in the estimated atmospheric parameters. However, a notable distinction arises in the case of the $u$-band. To investigate whether this dissimilarity could be attributed to an offset in the ZPs, we systematically applied a range of offsets for this band ($\delta u$), varying from $-0.09$ to $+0.4$ and repeating the process of SED fitting. Subsequently, we computed the disparity between the parameters obtained with the corrected magnitudes and those derived using the SDSS $u$-band.

\begin{figure}
\begin{center}
\includegraphics[width=0.47\textwidth]{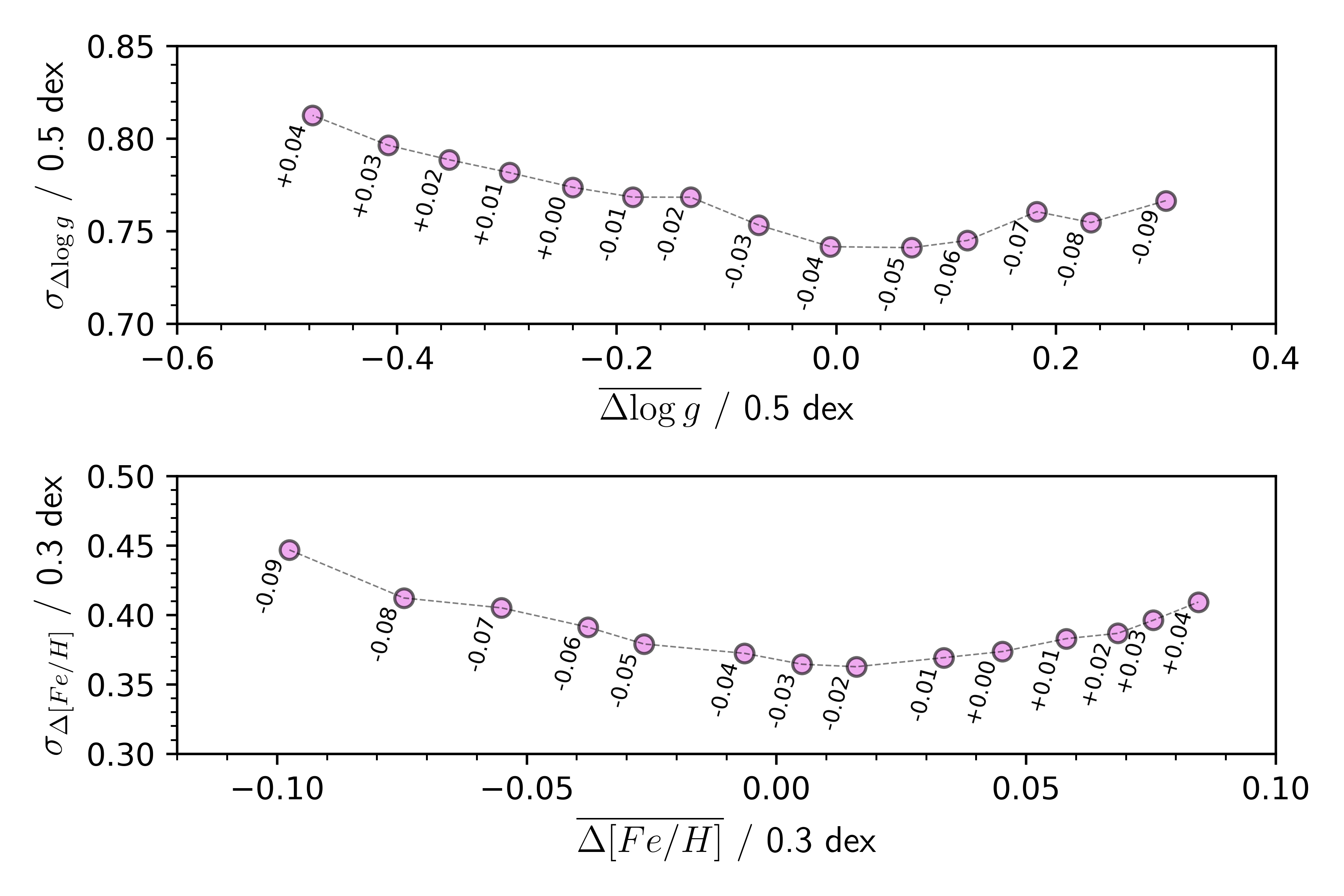}
\caption{Standard deviation versus offset (both normalized by their respective step in the model grid) of the difference between the fitted atmospheric parameters in each case (with $u' = u + \delta u$) and the ones fitted when substituting the $u$-band for $u$ from SDSS. The top panel shows the effects on $\log{g}$ and the lower panel on [Fe/H]. Each point represents a different $\delta u$ value, indicated on the corresponding label. A value of $-0.04$ mags is an appropriate correction for $u$ that would put the parameters closer to those fitted using SDSS.}
\label{fig:uoffset}
\end{center}
\end{figure}

Fig.~\ref{fig:uoffset} shows the normalized standard deviation and the offset of the differences between the parameters obtained in the case where we use $u+\delta u$ and the case using SDSS $u$-band as an input for the template fitting, which we adopt as reference values. For the $\log{g}$ (top panel), we found that a bias from $-0.04$ to $-0.05$ mags minimizes the difference and the standard deviation to the reference result. In the case of [Fe/H], a bias in the range $-0.03$ to $-0.04$ produces the best result. We thus recommend adding $-0.04$ mags to $u$ to correct it to the AB system.

The cause of this offset is outside of the scope of this article and still needs further thorough evaluation. It is important to mention that 40 mmag is still within the uncertainties associated with the $u$-band in this data release.

% Catalogs %%
\section{Catalogues}
\label{sec:catalogues}
%----------------------------------------------------------------

In this section, we present the structure of the data available in DR4, the \texttt{CLASS\_STAR} parameter, and the description of the value-added catalogues (VACs). Along with these, we also provide some useful information regarding the best practices suggested when using the catalogues.

%----------------------------------------------------------------
\subsection{Data structure}
%----------------------------------------------------------------

The data structure is similar to that of the former data releases, with the addition of the \single mode and the \PSF photometries. The DR4 is divided into four schemas: i) DR4\_dual contains tables related to the \dual mode photometry; ii) DR4\_single contains tables related to the \single mode; iii) DR4\_psf contains tables related to the \PSF; iv) DR4\_vacs contains the value-added catalogue tables, and will be expanded as the VACs are produced for DR4.
Appendix \ref{apd:columns} includes a brief description of each table's columns. Further information can be found in the S-PLUS documentation\footnote{\url{https://splus.cloud/documentation}}. 

The standard \texttt{SExtractor} photometric \texttt{FLAGS} are included in the catalogues. These flags are renamed \texttt{SEX\_FLAGS\_DET} for the \dual mode photometry and \texttt{SEX\_FLAGS\_\{filter\}} for \single mode. This bit-flag indicates several issues that can occur during photometry estimation, and the meaning of each bit can be found in the \texttt{SExtractor} manual. In addition, we also include another bit flag produced during the visual inspection of the calibrated catalogues. The listed value for a particular object corresponds to the sum of all flags occurring for the respective observation tile. Each bit represents the following:

\begin{itemize}
    \item[0 -] All good.
    \item[1 -] Unusual offset between \PSF and \dual modes magnitudes in at least one filter.
    \item[2 -] Larger than usual scatter between \PSF and \dual mode magnitudes in at least one filter.
    \item[4 -] Large offset between the field and the reference stellar locus.
    \item[8 -] Larger than usual scatter in the stellar locus.
    \item[16 -] Unreliable \texttt{CLASS\_STAR} and/or FWHM.
    \item[32 -] Correlation between the ZP offset and the magnitude in at least one filter.
    \item[64 -] Other.
\end{itemize}

%----------------------------------------------------------------
\subsection{The \texttt{CLASS\_STAR} parameter}
\label{sec::class_star}
%----------------------------------------------------------------

For the aperture photometry, \texttt{SExtractor} produces a \texttt{CLASS\_STAR} probability that ranges between 0 and 1 and indicates if the source behaves as a point source. However, this feature is not available in \texttt{DoPHOT}. In this case, the \PSF catalogue provides a star-galaxy classification obtained by an arbitrary selection considering the FWHM of the images. We made sure to be very inclusive in classifying the stars, only because we preferred to include misclassified sources instead of losing them. This value uses the same limits as the \texttt{SExtractor} \texttt{CLASS\_STAR} but, in this case, is a binary instead of a range, where 1 classifies the object as star and 0 otherwise.

\begin{figure*}
\includegraphics[width=\textwidth]{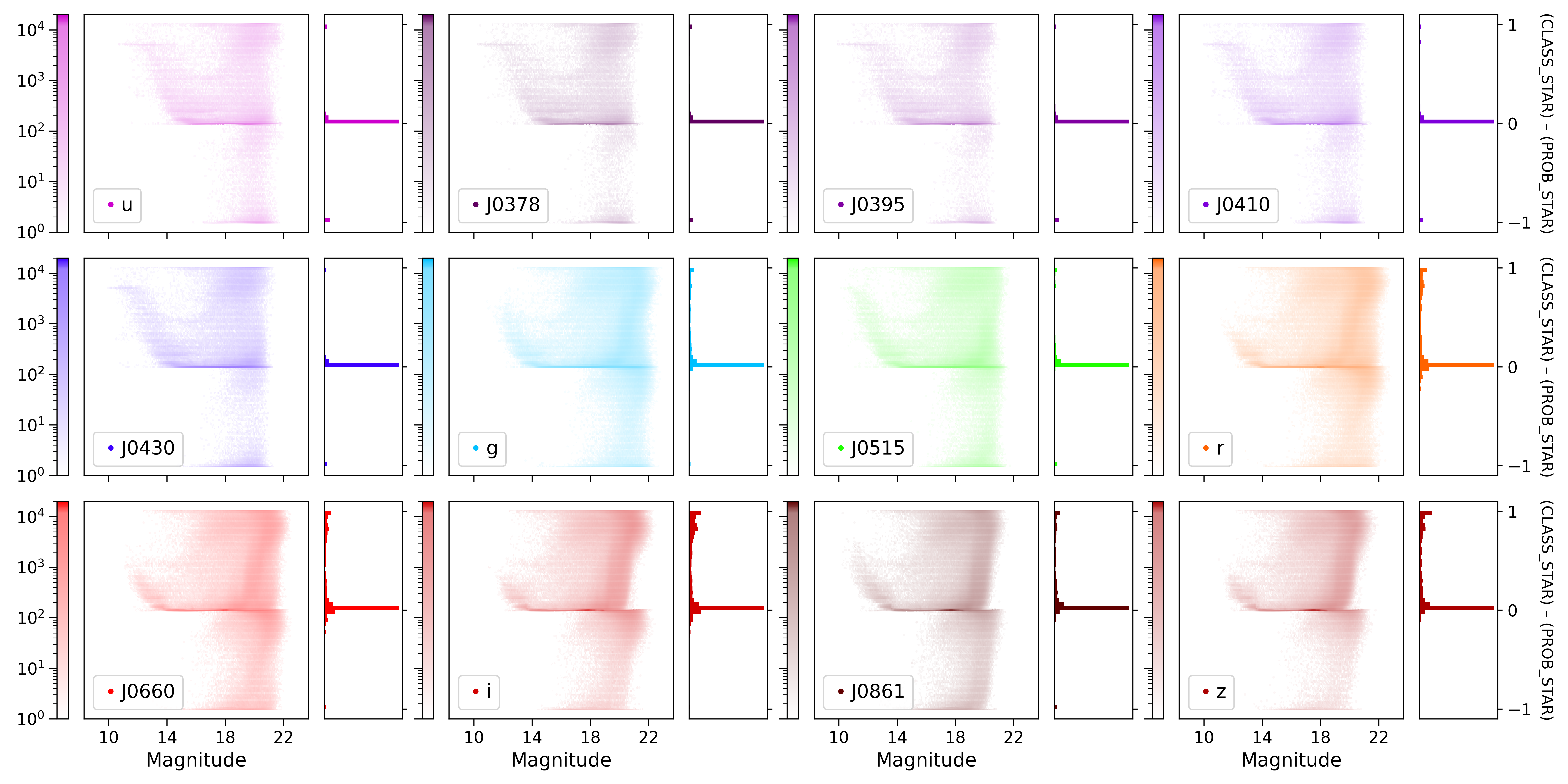}
\caption{Consistency between the classification from the \PSF photometry catalogue (\texttt{CLASS\_STAR}) and the probability of the source being a star from the star-galaxy-quasar classification VAC (\texttt{PROB\_STAR}) for each filter in the STRIPE82 region. The scatter plots correspond to the difference between the two classifications as a function of the magnitude, coloured by the logarithmic density of points. The correspondent histograms are normalized.}
\label{fig:psf_class}
\end{figure*}

This classification is compatible with the one found by the star-galaxy-quasar classification (see Section~\ref{sec:sgqsep}) for all 12 filters, with the differences between the two classifications as a function of the magnitude shown by Fig.~\ref{fig:psf_class}. We can see that, although both classifications are the same for most sources, they tend to differ from each other for fainter objects. In fact, only 13\% of the sources with $r < 20$ are misclassified by the \texttt{CLASS\_STAR} (by misclassification we mean objects with $\texttt{CLASS\_STAR} - \texttt{PROB\_STAR} > 0.5$ or $\texttt{CLASS\_STAR} - \texttt{PROB\_STAR} < -0.5$), where \texttt{PROB\_STAR} comes from the star-galaxy-quasar classification.

%----------------------------------------------------------------
%----------------------------------------------------------------

Moreover, the difference between \texttt{CLASS\_STAR} and \texttt{PROB\_STAR} is near 1 when we classify our sources using the \PSF photometry. Still, they have a low probability of being stars in the star-galaxy-quasar approach. The opposite condition happens less often, meaning fewer stars are misclassified as galaxies by the \PSF flag, where $\texttt{CLASS\_STAR} - \texttt{PROB\_STAR} \sim -1$. The latter is expected by construction, given that the \PSF class was tuned to avoid losing stars.

%----------------------------------------------------------------
\subsection{Value-added catalogues}
\label{sec:VACs}
%----------------------------------------------------------------
\subsubsection{Star-galaxy-quasar classification}\label{sec:sgqsep}
%----------------------------------------------------------------

Photometric classification of stars, galaxies, and quasars is provided for all sources in DR4, including the estimated probabilities for each class. The classification is performed by two Random Forest algorithms that were trained with \dual-photometric information of spectroscopically confirmed objects by SDSS DR16. For crowded fields, one should rely on \texttt{CLASS\_STAR} for a star-galaxy classification, as previously described in Section~\ref{sec::class_star}.

The first model is trained with 12 S-PLUS \iso magnitudes and four morphological features (\texttt{FWHM\_n}, \texttt{A}, \texttt{B}, \texttt{KRON\_RADIUS}). The second model includes the same features plus $W1$ and $W2$ magnitudes from AllWISE \citep{2013wise.rept....1C}, which constrains the dataset but increases the performance. We deliver the class estimated by the first model if a source does not have a WISE \citep{Wright+2010} counterpart (in either $W1$ or $W2$ bands). A column named \texttt{model\_flag} indicates which model was used, for which 1 means that the first model was considered, and 0 otherwise. Full details about data pre-processing and training strategy can be found in \citet{Nakazono+21}, as we kept the same strategy of DR2 and DR3. The columns provided by the DR4 star-galaxy-quasar VAC are listed in Table~\ref{tab:qso_class}, and the updated performance is shown in Table \ref{tab:sqg}. 

\begin{table}
\caption{Classification performance for DR4 in terms of the precision (P), recall (R), and F-score (F). It is important to note that the performances for the ``S-PLUS only'' feature space were calculated in a testing set containing only sources that do not have a WISE counterpart, i.e. most are at fainter sources, hence the decreased performance.}
\label{tab:sqg}
\resizebox{\columnwidth}{!}{%
\begin{tabular}{@{}llllll@{}}
\toprule
\textbf{Feature space} & \textbf{CLASS} & \textbf{P (\%)} & \textbf{R (\%)} & \textbf{F (\%)} \\ \midrule
S-PLUS + AllWISE                         & QSO            & 95.83                   & 96.46                & 96.14           \\
S-PLUS + AllWISE                          & STAR           & 99.50                   & 98.60                & 99.05           \\
S-PLUS + AllWISE                        & GAL            & 98.41                   & 98.96                & 98.68           \\
S-PLUS only                             & QSO            & 67.03                   & 93.02                & 77.92           \\
S-PLUS only                             & STAR           & 96.16                   & 76.70                & 85.34           \\
S-PLUS only                           & GAL            & 83.45                   & 86.32                & 84.86           \\ \bottomrule
\end{tabular}%
}
\end{table}

%----------------------------------------------------------------
\subsubsection{Photometric redshifts for galaxies}
\label{sec:photozgal}
%----------------------------------------------------------------

S-PLUS provides the single-point estimates (SPEs) of the photometric redshifts and probability distribution functions (PDFs) based on galaxy spectroscopy for all objects. These photometric redshift estimates are obtained using a Bayesian Mixture Density Network model \citep{Bishop94, Bishop97}, which is a supervised machine learning algorithm. More details can be found in \citet{Lima+22}.

For the supervised learning models, we assembled a training sample containing input properties and a target. The input features combine photometric features (magnitudes and colours in the \texttt{aper\_6 aperture}) and non-photometric features (detection radius). The photometry from S-PLUS is complemented by $J$, $H$, and $K$ magnitudes from the 2MASS \citep{Skrutskie+2006}, and $W1$ and $W2$ magnitudes from the unWISE \citep{Lang14} catalogues. The model aims to predict the redshift of galaxies based on a compilation of spectroscopic catalogues (Lima et al., in prep.). 

The final training sample contains 299\,602 galaxies after pre-processing. This sample is split into training, validation, and testing sets, with 81\%, 9\%, and 10\% objects, respectively. The performance metrics reported in this work are calculated using the testing set.

The model is trained on a sample containing only galaxies with $14 \leq r \leq 21.3$ mag and respective errors below $0.5$ mag. Any estimate made for objects beyond these ranges may be considered extrapolation, and caution is advised. If an estimate for such objects is desired, we recommend evaluating the value of the odds for that prediction, which is the quantity that represents the area of the probability distribution function inside the interval $z_\text{phot} \pm 0.002$, for which we suggest a value of $0.4$ or higher to be used.

\begin{figure*}
    \centering
    \includegraphics[width=\textwidth]{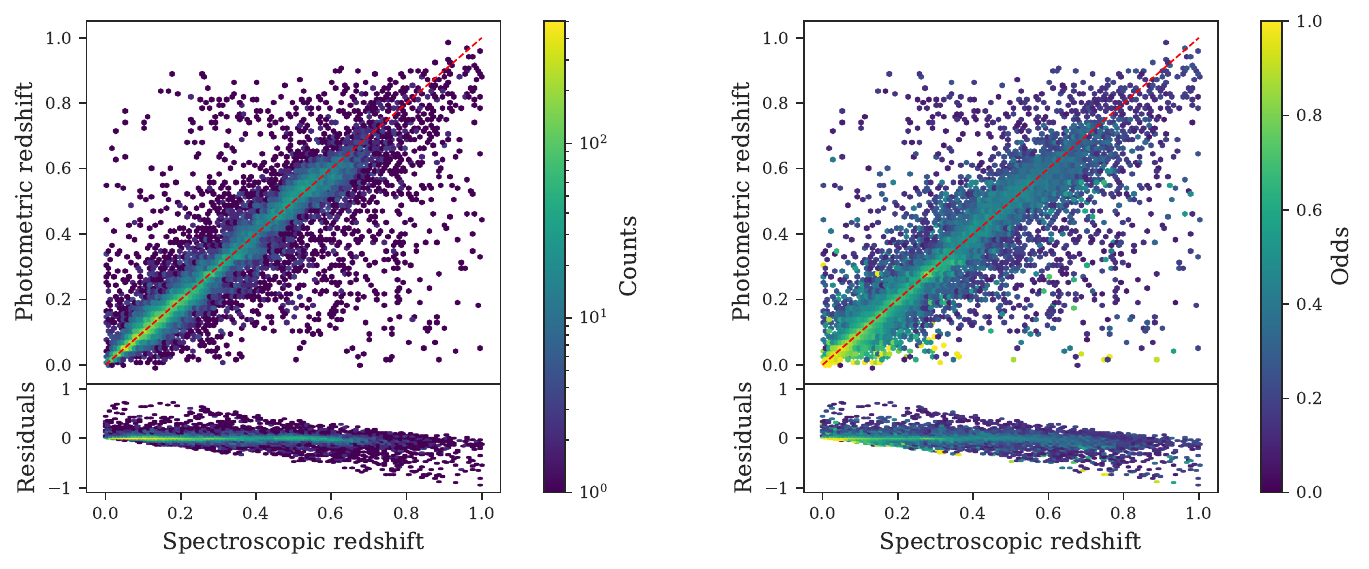}
    \caption{Number count density (left) and odds (right) of spectroscopic versus photometric redshifts for the testing sample. The residuals (bias) are represented below their respective plots.}
    \label{fig:photoz_scatter}
\end{figure*}

The results of the testing sample show that our model can provide accurate and low-biased single-point estimates while also generating well-calibrated probability distribution functions, which can reliably account for the uncertainties in the estimates. Fig.~\ref{fig:photoz_scatter} shows the prediction scatter coloured by number density in logarithmic scale and by the value of the odds \citep{Benitez2000}. The metrics used and the performance calculated for the single-point estimates and probability density functions are presented in Appendix \ref{apd:photoz_metrics}.

%----------------------------------------------------------------
%----------------------------------------------------------------

%When analyzing the bias as a function of redshift we see that, for spectroscopic redshifts below 0.007, the bias is of the same order as the redshift measurement itself, and this is the reason for the recommendation.

Our analysis also shows that the photometric redshifts should not be used for objects with redshifts below $0.007$ or above $0.6$. Both limits are defined due to the bias in the estimates, defined in Eq. \eqref{eq:bias} and calculated using a testing sample. For the lower end, the bias is of the same order as the redshift measurement itself, while for the upper end, we see an increase in this metric (as shown in panel (b) of Fig. \ref{fig:spe_metrics}). Both situations arise due to the lack of training data in these regions of the featured space. Future S-PLUS data releases, which will cover a larger area in the sky, will improve the model's performance in these cases due to the larger quantity of overlapping objects, which leads to a larger training sample for the model.

%----------------------------------------------------------------
\subsubsection{Photometric redshifts for quasars}\label{sec:photquasars}
%----------------------------------------------------------------

The photometric redshifts focusing on quasars are also available for the DR4. We use three different machine learning codes, namely Random Forest \citep{Breiman2001}, Bayesian Mixture Density Networks (based on \citealt{Lima+22}; see Section~\ref{sec:photozgal}), and FlexCoDE \citep{izbicki}. These models are trained on a spectroscopic dataset from the SDSS DR16 Quasar Catalogue \citep{Lyke+20}. More details can be found in \citet{2024MNRAS.531..327N}. In DR4, we provide the SPEs for all methods and the PDFs for the Bayesian Mixture Density Networks and FlexCoDE. The provided columns are fully described in Table \ref{tab:qso_z}.

%----------------------------------------------------------------
\subsubsection{Stellar parameters using CNN}

This catalogue provides effective temperature ($T_\mathrm{eff}$), metallicity ([Fe/H]), and surface gravity ($\log g$) for approximately 3.5 million stars in S-PLUS DR4. The parameters were determined using a one-dimensional convolutional neural network (CNN) architecture. The CNN was trained using the 12-band photometry data from S-PLUS as features and $T_\mathrm{eff}$, [Fe/H], and ($\log g$) from LAMOST DR8 as labels. After the application of quality cuts, the common sample between these catalogues consists of around 55\,000 stars, of which 80\% were used for training and 20\% for testing. The mean differences between the values estimated by the CNN and the LAMOST data for the test sample are approximately zero, and the $\sigma$-scatter are around 115 K for $T_\text{eff}$, and 0.2 dex for both [Fe/H] and $\log g$. Additional details regarding the CNN architecture, application intervals, and quality cut-offs applied to both the training and test data can be found in \citet{2024MNRAS.527.6173Q}. The columns available in this VAC are presented in Table \ref{tab:VAC_CNN}.

\subsubsection{Stellar parameters using SPHINX}

We also obtained stellar parameters for S-PLUS DR4 by applying the same technique described in the work of \citet{2021ApJ...912..147W} for the S-PLUS DR2. We only considered sources classified as stars by the star-galaxy-quasar classification ($\texttt{CLASS} = 1$), with a detection \SN of $\texttt{s2n\_DET\_auto} \geq 3$, and \texttt{PSTotal} measurements in all 12 S-PLUS filters. The parameters are obtained using the Stellar Photometric Index Network Explorer (\texttt{SPHINX}). \texttt{SPHINX} is an algorithm based on a consensus procedure for a set of Artificial Neural Networks (ANN), in which each ANN sub-unit is fed with a unique combination of S-PLUS colours. A more complete description of \texttt{SPHINX} is presented in \citet{2019A&A...622A.182W}.

This dataset provides effective temperature, $T_\mathrm{eff}$, metallicity, [Fe/H], and absolute carbon abundance, A(C)\footnote{A(C) = $\log(n_C / n_H) + 12$, where $n_C$ and $n_H$ represents the number densities of Carbon and Hydrogen species, respectively.}, for approximately 4.3 million stars. The catalogue also contains estimated errors and the number of ANN sub-units used for the estimation of each parameter. As described in \citet{2021ApJ...912..147W}, different ANN structures are used for [Fe/H] and A(C) depending on the temperature range ([4250, 5750] K, and [5500, 7000] K). For stars within the overlap region of the two sets ([5500, 5750] K) the reported value corresponds to the average of the two estimates (these cases are indicated by a flag included in these data). A description of each column in this VAC is presented in Table \ref{tab:VAC_ANN}.

%----------------------------------------------------------------
\subsubsection{Best practices with VACs}
%----------------------------------------------------------------

Each VAC has its limitations due to the chosen algorithm and its respective training set. Here, we provide some recommendations on how to define selection criteria that align with the algorithm's expected performance.

For the star-galaxy-quasar classification, we recommend using only sources with $\texttt{SEX\_FLAGS\_r}=0$ and $13 \leq r \leq 21.3$ mag with a respective error of $0.2$ mag or less. For the photometric redshifts it is important to note that, even though this VAC is available for all objects independent of their classification, the training sample only consists of galaxies, thus redshifts for objects classified as stars must be ignored. Predictions for quasars might not be sufficiently reliable and we recommend the reader to query the quasars VAC instead (see Section~\ref{sec:photquasars}).

We also recommend using the FlexCoDE SPEs and PDFs for quasars, given that they provide the lowest loss function and precision of $\sigma_{\textrm{NMAD\footnotemark}}=0.042$\footnotetext{Normalized Mean Absolute Deviation.}. Most confident predictions are for sources within $17.5 \lessapprox r  \lessapprox 19.5$, with a precision up to $0.02$. Sources beyond the $r$ depth should be used with caution, as the accuracy ranges from $0.06$ to $0.12$.

% Data Acess %%
\section{Data access}
\label{sec:data_access}
%----------------------------------------------------------------

%----------------------------------------------------------------
\subsection{The splus.cloud database: Data and image access tools}
%----------------------------------------------------------------

All S-PLUS data, including previous data releases, are hosted in the collaboration server splus.cloud\footnote{\url{https://splus.cloud/}}. In the Tools section of the website, the reader can find several tools for different tasks, as follows:

\begin{itemize}
    \item \texttt{Image Tools > 1 Filter Image}: Download a PNG cut-out of a single filter.
    \item \texttt{Image Tools > 3 Filters Image}: Download a coloured PNG cut-out by choosing one filter for each channel.
    \item \texttt{Image Tools > 12 Filters Image}: Download a coloured PNG cut-out by choosing a combination of filters for each channel.
    \item \texttt{Image Tools > FITS image}: Download a FITS image cut-out for a selected filter.
    \item \texttt{Image Tools > Field FITS image}: Download the $11000x11000$ pixels FITS coadded image of a given field for a selected filter. Weight images are also available.
    \item \texttt{Catalogue Tools > Check Coordinate}: Check if a pair of coordinates is present in this or previous data releases.
\end{itemize}

In addition, splus.cloud integrates the Table Access Protocol (TAP), conforming to the standards set by the International Virtual Observatory Alliance (IVOA), described in \cite{IVOA_TAP}. TAP facilitates efficient access to the database, enabling users to execute adhoc queries in the SQL-like Astronomical Data Query Language (ADQL) against the data tables stored in the database. This feature is particularly useful for researchers requiring customized data retrieval and analysis. A walk-through to access the data is available at DR4 documentation page\footnote{\url{https://splus.cloud/documentation/DR4?Catalogs}}. 

%----------------------------------------------------------------
\subsection{The \texttt{splusdata} python package}
%----------------------------------------------------------------

The \texttt{splusdata} Python module provides integration with splus.cloud, enabling direct data access and manipulation within Python. This module supports automated data retrieval and processing, which is important to researchers needing streamlined workflows and advanced data analysis capabilities, including machine learning applications. Installation options include \texttt{pip} and direct download from GitHub\footnote{\url{https://github.com/Schwarzam/splusdata}}.

% Science Cases%%
\section{The feasibility of the identification of galaxy cluster membership using S-PLUS DR4}
\label{sec:science_cases}

As a case test, we used a DR4 sample selected from the splus.cloud to identify galaxy cluster members in the low- and mid-redshift regimes ($0.05 \lesssim z \lesssim 0.3$, respectively), showcasing the versatility of the S-PLUS data. Our approach involved selecting four Abell clusters (1300, 3158, 3880, and 1644) and obtaining their photometry and photometric redshifts from the splus.cloud database using a cone-search centred at their right ascension and declinations with a 1.5 deg radius for the first cluster, and $5 \times r_{200}$ for the others.

\begin{figure*}
    \centering
    \includegraphics[width=\textwidth]{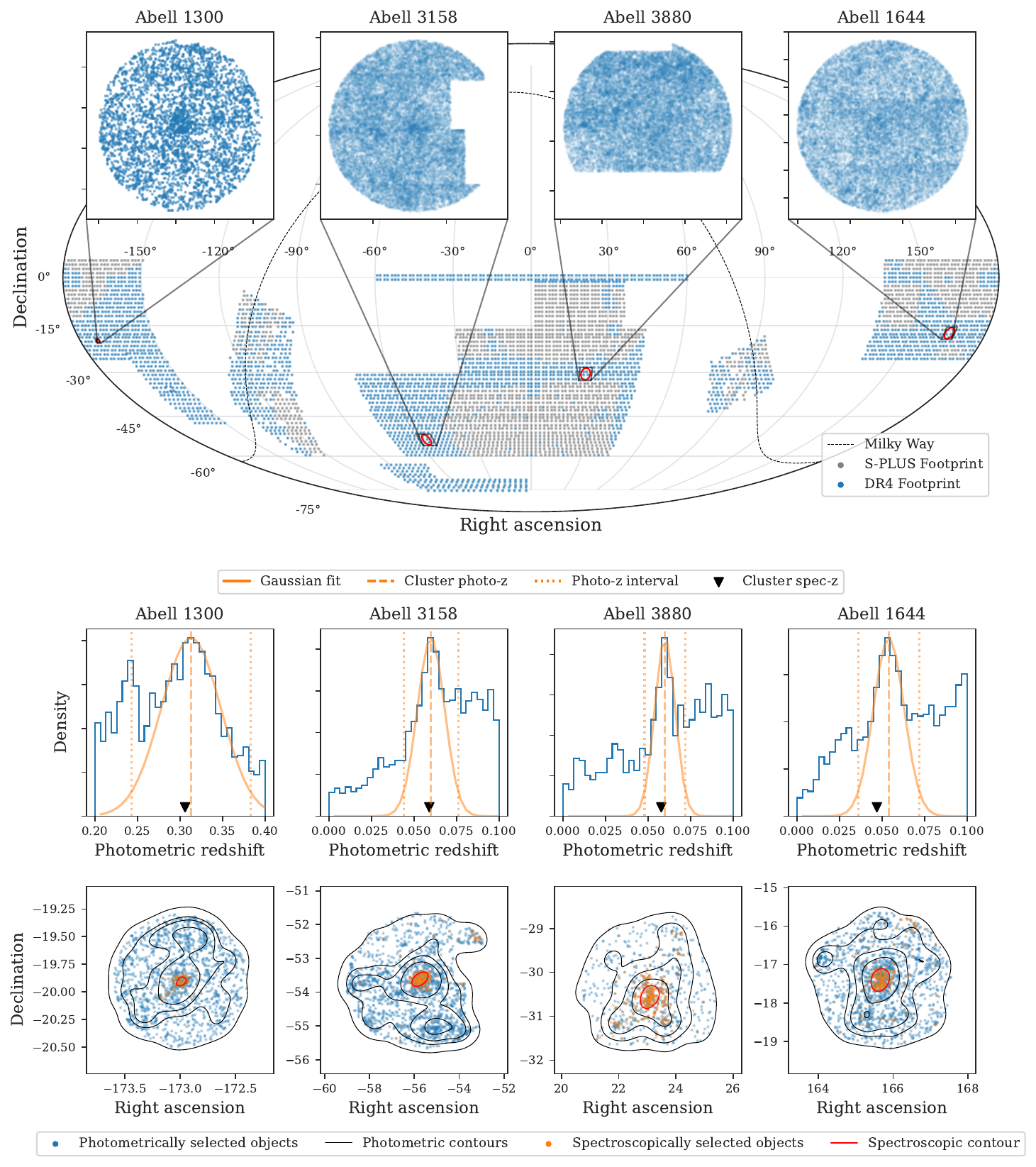}
    \caption{Identification of cluster membership with S-PLUS photometric redshifts. The top row contains the sky projection of the S-PLUS data galaxies ($\texttt{PROB\_GAL} \geq 0.5$) obtained from a cone-search centred in the cluster coordinates. The middle row contains the photometric redshift distribution for these clusters, and the bottom row shows the objects selected based on a cut on photometric redshifts.}
    \label{fig:photoz_clusters}
\end{figure*}

After downloading the data, we selected only objects with $\texttt{PROB\_GAL} \geq 0.5$ and \texttt{r\_aper\_6} $\leq 21$, and created a histogram of the photometric redshifts, shown by Fig.~\ref{fig:photoz_clusters}. In this plot, we identify a peak in the distribution close to the expected cluster redshifts. A fit of a single Gaussian distribution to these peaks allows us to find the mean and standard deviations of the cluster's photometric redshifts, which we then used to select all objects within the $2\sigma$ range. The spectroscopic and mean photometric redshifts and standard deviations obtained from the Gaussian fit for each cluster are listed in Table \ref{tab:photoz_clusters}. With this simplistic approach, we can select a sample of probable cluster members that spans to bigger radii and a larger number of members than those available in the current spectroscopic samples. The relevance of this kind of application is to provide a larger sample of candidates belonging to a cluster in outskirts regions (up to $5 \times r_{200}$), facilitating the study of the environment and its relation to the galaxies therein.

\begin{table}
    \caption{Spectroscopic and mean photometric redshifts with their respective standard deviations obtained with a Gaussian fit for the four clusters studied in this work: Abell 1300, Abell 3158, Abell 3880, and Abell 1644.}
    \label{tab:photoz_clusters}
    \centering
    \begin{tabular}{@{}lcc@{}}
        \toprule
        Cluster & Spectroscopic redshift & Photometric redshift \\
        \midrule
        Abell 1300 & $0.304 \pm 0.009$ & $0.313 \pm 0.035$ \\
        Abell 3158 & $0.060 \pm 0.004$ & $0.060 \pm 0.008$ \\
        Abell 3880 & $0.057 \pm 0.003$ & $0.060 \pm 0.007$ \\
        Abell 1644 & $0.047 \pm 0.004$ & $0.054 \pm 0.009$ \\
        \bottomrule
    \end{tabular}
\end{table}

% Summary %%
\section{Summary}
\label{sec:summary}
%----------------------------------------------------------------

In this work, we presented the S-PLUS's fourth data release, which includes sky-subtracted images and catalogues with photometric characterization and calibrated AB magnitudes in 12 filters (seven narrow and five broadbands). These data cover 3022.7 sq deg in the southern sky. This area includes and expands the regions covered in the previous data releases along with new areas introduced in the DR4, namely fields with low galactic latitude in the anti-centre direction of the Galaxy and the Magellanic Clouds.

This is also the first S-PLUS data release to include PSF photometry, which is the optimal option for the crowded regions introduced in this dataset. In addition to the dual-mode photometry presented in the previous releases, DR4 includes single-mode aperture photometry, where each filter's measurements and detections are computed independently. We employed the same photometric calibration technique used in the previous releases, with improvements in treating the ISM extinction during the calibration process. We note, however, that the published magnitudes are not corrected by ISM extinction. 

Different calibration strategies (in terms of reference catalogues, base instrumental photometry, and selection of ISM extinction map) were required for the calibration of different regions of the survey. We applied these strategies to the same fields in the STRIPE82 region to estimate the internal errors among the methods. For the strategies used in the MS, we find that, in general, the expected errors are $\lessapprox 20$ mmags for the blue filters ($u$, $J0378$, $J0395$), and $\lessapprox 10$ mmags for the others. The errors are slightly higher for the regions of the disk and MC, where the ISM extinction is larger. In the case of the Disk, the internal errors of the ZPs are $\approx 30$ mmags for the blue filters ($u$, $J0378$, $J0395$) and $\lessapprox 15$ mmags for the remaining. These uncertainties are similar to those found in the calibration of the MC, except when the calibration is based on SkyMapper DR1.1 \citep[][]{Wolf+2018}\footnote{SkyMapper DR1.1 Catalogue DOI: 10.4225/41/593620ad5b574.}, which results in slightly larger uncertainties: $\lessapprox 60$ mmags (for $u$, $J0378$, $J0395$), $\lessapprox 35$ mmags (for $J0410$ and $J0430$) and $\lessapprox 15$ mmags for all others.

The absolute error of the calibration was estimated by converting S-PLUS photometry to the Gaia DR2 system and comparing the results to the published Gaia magnitudes. The average offsets are $0.017 \pm 0.006$, $-0.040 \pm 0.007$, and $-0.059 \pm 0.014$ for the MS, MC and disk, respectively. These offsets are already adhoc corrected in the reported magnitudes.

The photometric depths of all MS fields and bands were calculated using the \petro magnitudes. Each field had a different depth calculated by the peak of the magnitude distribution for all 12 bands for four different $S/N$ ratios: 3, 5, 10, and 50. Then, we accounted for the depths of all fields to characterize the whole MS. For a $S/N$ threshold of five, we had a minimum depth of $\sim 19$ mag for the $J0395$-band and a maximum of $\sim 20.4$ mag for the $r$-band. The typical dispersion in depths is around $0.4$ mag, with higher dispersion for bluer bands (e.g. $\sim0.4$ mag for $J0430$) and smaller for redder bands (e.g. $\sim0.2$ mag for the $z$-band).

As an application of DR4, we have shown that we can identify additional members of the clusters Abell 1300, Abell 1644, Abell 3158, and Abell 3880 through an analysis of the photometric redshifts in the line of sight of these systems, allowing us to measure their redshifts (0.313, 0.054, 0.060, and 0.060, respectively). The main advantage of this approach is the use of photometric redshifts obtained for a large number of galaxies in an extended region, not limited to the inner parts of the clusters (as usually is the case in spectroscopic studies). This result shows that S-PLUS can be used to identify and characterize new galaxy clusters, considering the environment's properties, while also expanding the cluster members away to distances far from the centre, enabling studies about their outskirts.

S-PLUS is soon entering its 8th year of operation (science operations started in August 2017, \citetalias{MendesDeOliveira+19}) with over 70\%\ of the originally planned footprint already observed (with about a third of the area becoming public now). The availability of such a large dataset will deliver on the promises outlined in \citetalias{MendesDeOliveira+19} on the possible synergies, such as with the Gaia mission and other ongoing and upcoming photometric and spectroscopic wide-area surveys, such as SDSS V \citep{2023ApJS..267...44A}, DESI \citep{2024AJ....168...58D}, Euclid \citep{2022AA...662A.112E}, LSST, and Roman (previously known as WFIRST, \citealt{2015`150303757S}). The majority of the southern 10- and 30-meter class photometric and spectroscopic facilities will have a substantial overlap with the planned S-PLUS footprint, thus holding promising prospects in the upcoming years.

%% Acknowledgements %%
\section*{Acknowledgements}

The S-PLUS project, including the T80-South robotic telescope and the S-PLUS scientific survey, was founded as a partnership between the Fundação de Amparo à Pesquisa do Estado de São Paulo (FAPESP), the Observatório Nacional (ON), the Federal University of Sergipe (UFS), and the Federal University of Santa Catarina (UFSC), with important financial and practical contributions from other collaborating institutes in Brazil, Chile (Universidad de La Serena), and Spain (Centro de Estudios de Física del Cosmos de Aragón, CEFCA). We further acknowledge financial support from the São Paulo Research Foundation (FAPESP) grant 2019/263492-3, the Brazilian National Research Council (CNPq), the Coordination for the Improvement of Higher Education Personnel (CAPES), the Carlos Chagas Filho Rio de Janeiro State Research Foundation (FAPERJ), and the Brazilian Innovation Agency (FINEP).

The S-PLUS collaboration members are grateful for the contributions from CTIO staff in helping in the construction, commissioning, and maintenance of the T80-South telescope and camera. We are also indebted to Rene Laporte, INPE, and Keith Taylor for their essential contributions to the project. From CEFCA, we particularly would like to thank Antonio Marín-Franch for his invaluable contributions in the early phases of the project, David Cristóbal-Hornillos and his team for their help with the installation of the data reduction package JYPE version 0.9.9, César Íñiguez for providing 2D measurements of the filter transmissions, and all other staff members for their support with various aspects of the project.

F.R.H. acknowledges support from FAPESP grants 2018/21661-9 and 2021/11345-5.
F.A.-F. acknowledges funding for this work from FAPESP grants 2018/20977-2 and 2021/09468-1.
L. N. acknowledges the support from FAPESP (grants 2019/01312-2 and 2021/12744-0)
J.A.-G. acknowledges support from Fondecyt Regular 1201490 and from ANID – Millennium Science Initiative Program – ICN12\_009 awarded to the Millennium Institute of Astrophysics MAS.
M.J.S. acknowledges financial support from grant 2022/00996-8, São Paulo Research Foundation (FAPESP).
A.W. has received funding from the European Research Council (ERC) under the European Union’s Horizon 2020 research and innovation program (grant agreement No. 833824, GASP project).
E. M.-P. acknowledges CAPES (proc. 88887.605761/2021-00).
L.A.G.S. and A.V.S.C. acknowledge financial support from CONICET, Agencia I+D+i (PICT 2019-03299) and Universidad Nacional de La Plata (Argentina).
CEFL thanks for the support of ANID's Millennium Science Initiative through grant ICN12\_12009, awarded to the Millennium Institute of Astrophysics (MAS), and ANID/FONDECYT Regular grant 1231637.
M.S.C. acknowledges support from São Paulo Research Foundation (FAPESP), grant 2023/10774-5.
P.C. acknowledges support from Conselho Nacional de Desenvolvimento Cient\'ifico e Tecnol\'ogico (CNPq) under grant 310555/2021-3 and from Funda\c{c}\~{a}o de Amparo \`{a} Pesquisa do Estado de S\~{a}o Paulo (FAPESP) process number 2021/08813-7.
L.D. acknowledges the support from the scholarship from the Brazilian federal funding agency Coordenação de Aperfeiçoamento de Pessoal de Nível Superior - Brasil (CAPES) and FAPESP (\#2024/03575-9).
F.Q.H. acknowledges financial support from CNPq Grant 306009/2019-6.
G.J.P. acknowledges support from FAPESP (grant 2022/11645-1).
L. L.-N. thanks Funda\c{c}\~ao de Amparo \`a Pesquisa do Estado do Rio de Janeiro (FAPERJ) for granting the postdoctoral research fellowship E-40/2021(280692)
RLO acknowledges financial support from the Brazilian institutions CNPq (PQ-312705/2020-4) and FAPESP (\#2020/00457-4).
A.R.L. acknowledges financial support from Consejo Nacional de Investigaciones Científicas y Técnicas (CONICET) and Agencia I+D+i (PICT 2019–03299) (Argentina).
The work of FN is supported by NOIRLab, which is managed by the Association of Universities for Research in Astronomy (AURA) under a cooperative agreement with the National Science Foundation. AFJM is grateful for financial aid from NSERC (Canada).
The work of V.M.P. is supported by NOIRLab, which is managed by the Association of Universities for Research in Astronomy (AURA) under a cooperative agreement with the National Science Foundation.
VHS thanks CNPq for the financial support.
J.T-B. acknowledges financial support from FAPESC (CP 48/2021).
T.C.B. acknowledges partial support for this work from grant PHY 14-30152; Physics Frontier Center/JINA Center for the Evolution of the Elements (JINA-CEE), and OISE-1927130: The International Research Network for Nuclear Astrophysics (IReNA), awarded by the US National Science Foundation.
G.L. acknowledges FAPESP grant 2021/10429-0.
P.A.A.L. thanks the support from CNPq, grants 433938/2018-8 and 312460/2021-0 and FAPERJ, grant E- 26/200.545/2023.
G.P.M. acknowledges financial support from ANID/``Beca de Doctorado Nacional''/21202024.
L.BeS. acknowledges the support provided by the Heising Simons Foundation through the Barbara Pichardo Future Faculty Fellowship from grant \#2022-3927.
P.K.H. gratefully acknowledges the Fundação de Amparo à Pesquisa do Estado de São Paulo (FAPESP) for the support grant 2023/14272-4.
M.B.F. acknowledges financial support from the National Council for Scientific and Technological Development (CNPq) Brazil (grant number: 307711/2022-6).
V.C. acknowledges Coordenação de Aperfeiçoamento de Pessoal de Nível Superior - Brasil (CAPES) - Finance Code 001.

This work has made use of data from the European Space Agency (ESA) mission
Gaia\footnote{\url{https://www.cosmos.esa.int/gaia}}, processed by the Gaia
Data Processing and Analysis Consortium (DPAC\footnote{
\url{https://www.cosmos.esa.int/web/gaia/dpac/consortium}}. Funding for the DPAC
has been provided by national institutions, in particular, the institutions
participating in the Gaia Multilateral Agreement.

The authors thank Ulisses Manzo Castello, Marco Antonio dos Santos, and Luis Ricardo Manrique for supporting infrastructure matters.

%%%%%%%%%%%%%%%%%%%%%%%%%%%%%%%%%%%%%%%%%%%%%%%%%%
%% data availability %%
\section*{Data availability}

All the data in this article has been made publicly available as of Dec 18th, 2023, at the S-PLUS database splus.cloud\footnote{\url{https://splus.cloud/}}, which follows the standards defined for the Virtual Observatory environment and data accessibility.

\bibliographystyle{aa}
\bibliography{bibliography} % if your bibtex file is called example.bib

%%%%%%%%%%%%%%%%%%%%%%%%%%%%%%%%%%%%%%%%%%%%%%%%%%

%%%%%%%%%%%%%%%%% APPENDICES %%%%%%%%%%%%%%%%%%%%%
%% Appendiced %%
\appendix
% \input{tex/appendix-institutions}
%----------------------------------------------------------------
\section{The S-PLUS filter curves}
\label{ap:splusfilters}
%----------------------------------------------------------------

\begin{figure*}
\begin{center}
\includegraphics[width=1.95\columnwidth]{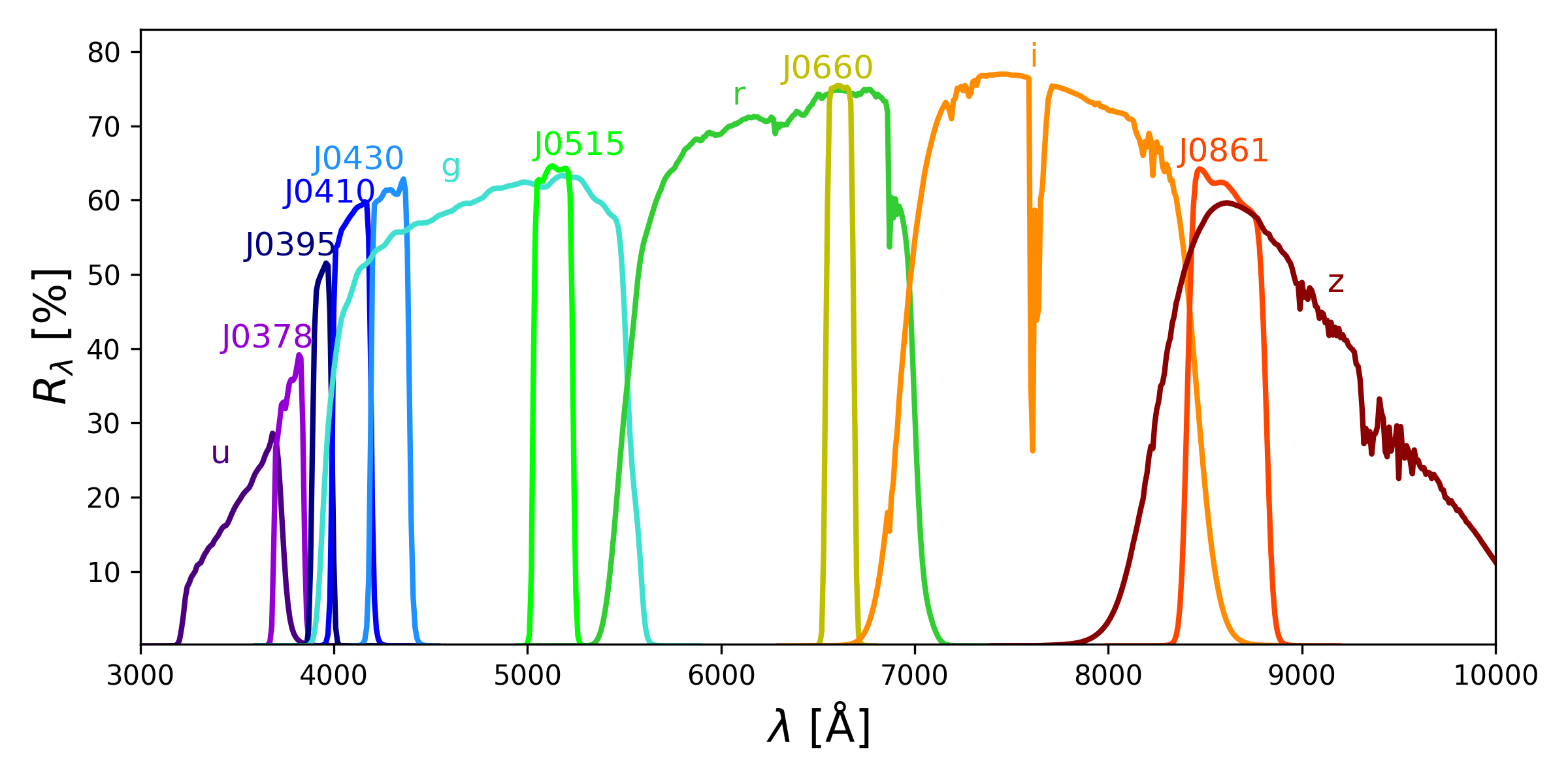}
\caption{S-PLUS convolved transmission curves after accounting for the instrumental and atmospheric contributions.}
\label{fig:splus_filters}
\end{center}
\end{figure*}

The S-PLUS project received the laboratory measures of the filter transmission from the Centro de Estudios de F\'isica del Cosmos de Arag\'on (CEFCA) in 2015. Using these measurements, we calculated the convolved curves after including the atmospheric transmission \citep{2012A&A...543A..92N}, the CCD efficiency measured by e2v\footnote{\url{https://www.teledyne-e2v.com/}}, and mirror reflectivity measured at the CTIO in 2016, which resulted in the curves used since then, which were published in \citetalias{MendesDeOliveira+19}. More recently, we encountered some inconsistencies between our measurements of central lambdas and those estimated by external sources using the same transmission curves, thus raising the need to revisit these calculations and make this an official release of the parameters related to the S-PLUS filters. In addition to calculating the central lambdas and the FWHM already published in \citetalias{MendesDeOliveira+19}, we calculated the lambda mean, mean width, lambda pivot, and the $A_\lambda/A_V$ using the extinction curves of \citet{1999PASP..111...63F}\footnote{The opacity spectrum from \citet{1999PASP..111...63F} and \citet{2005ApJ...619..931I} was obtained directly from the Spanish Virtual Observatory hosted at \url{https://svo.cab.inta-csic.es/main/index.php}.} and \citet{Cardelli+89-v345-p245}. The resulting transmission curves from this exercise are shown in Fig.~\ref{fig:splus_filters}.

Table~\ref{tab:central_wavelengths} lists the aforementioned measurements. All values were calculated over the convolved curves to reflect the reality of the filters. The $\lambda_{\mathrm{central}}$ corresponds to the mean point at half the curve's height. A linear interpolation was used to get the most accurate wavelength and decrease the dependence on the granularity of the data. The FWHM is exactly what its definition represents, the width at half the height. $\lambda_{\mathrm{mean}}$ and $\Delta\lambda_{\mathrm{mean}}$ are defined by Eq. \ref{eq:lambda_mean}.

\begin{table*}
\centering
\caption{Central wavelengths of the S-PLUS filters obtained via the different definitions described in this section.}
\label{tab:central_wavelengths}
\begin{tabular}{cccccccc}
\toprule
Filter & $\lambda_{\mathrm{central}}$ & FWHM & $\lambda_{\mathrm{mean}}$ & $\Delta\lambda_{\mathrm{mean}}$ & $\lambda_{\mathrm{pivot}}$ & $A_{\lambda}/A_{V}$ & $A_{\lambda}/A_{V}$\\
 & [\AA] & [\AA] & [\AA] & [\AA] & [\AA] & \citep{1999PASP..111...63F} & \citep{Cardelli+89-v345-p245} \\
\midrule
$u$ & 3577 & 325 & 3542 & 323 & 3533 & 1.610 & 1.584\\
$J0378$ & 3771 & 151 & 3774 & 136 & 3773 & 1.518 & 1.528\\
$J0395$ & 3941 & 103 & 3941 & 101 & 3941 & 1.459 & 1.483\\
$J0410$ & 4094 & 200 & 4096 & 193 & 4095 & 1.403 & 1.434\\
$J0430$ & 4292 & 200 & 4293 & 195 & 4292 & 1.334 & 1.364\\
$g$ & 4774 & 1505 & 4821 & 1312 & 4758 & 1.199 & 1.197\\
$J0515$ & 5133 & 207 & 5134 & 204 & 5133 & 1.098 & 1.085\\
$r$ & 6275 & 1437 & 6296 & 1274 & 6252 & 0.864 & 0.866\\
$J0660$ & 6614 & 147 & 6614 & 147 & 6614 & 0.798 & 0.810\\
$i$ & 7702 & 1507 & 7710 & 1438 & 7671 & 0.648 & 0.646\\
$J0861$ & 8611 & 410 & 8610 & 402 & 8607 & 0.539 & 0.518\\
$z$ & 8882 & 1270 & 8986 & 1308 & 8941 & 0.512 & 0.484\\
\bottomrule
\end{tabular}
\end{table*}

\begin{equation}\label{eq:lambda_mean}
\begin{split}
    \lambda_{\mathrm{mean}} & = \frac{\int \lambda T(\lambda) d\lambda}{\int T(\lambda) d\lambda}, \\
    \Delta\lambda_{\mathrm{mean}} & = \frac{\int T(\lambda) d\lambda}{Max\left(T(\lambda)\right)},
\end{split}
\end{equation}

\noindent
where $\lambda$ is the wavelength and $T(\lambda)$ is the transmission $T(\lambda)=\lambda R(\lambda)$ with $R(\lambda)$ being the filter response. The $\lambda_{\mathrm{pivot}}$ is calculated following Eq. \ref{eq:lambda_pivot}, and is the parameter used to estimate $A_\lambda / A_V$.

\begin{equation}
    \label{eq:lambda_pivot}
    \lambda_{\mathrm{pivot}} = \sqrt{\frac{\int T(\lambda) d\lambda}{\int \left(\frac{T(\lambda)}{\lambda^2}\right) d\lambda}}.
\end{equation}

All calculations, codes, and measurements used to define the filter transmissions are available at the S-PLUS collaboration GitHub page\footnote{\url{https://github.com/splus-collab/splus_filters}}.

\section{Photometric redshift performance metrics}
\label{apd:photoz_metrics}

To quantify the performance of the photometric redshifts, we selected two different sets of metrics: one for the probability distribution functions and another for the single-point estimations. For the probability distribution functions, it is recommended to analyse more than one metric to evaluate the quality of the PDFs \citep{Schmidt2020}. For this reason we calculate the Probability Integral Transform \citep[PIT, ][]{Polsterer+16},  Highest-Probability Density Credible Interval \citep[HPDCI, ][]{Soto2002}, Continuous Ranked Probability Score \citep[CRPS, ][]{Hersbach2000}, and the odds distribution \citep{Benitez2000}. These metrics are defined by Eq.~\ref{eq:PIT} to \ref{eq:Odds}.

\begin{align}
  \label{eq:PIT}
  \text{PIT} = \int_{0}^{z_\text{spec}} \text{PDF}_i(z)~\text{d}z, 
\end{align}
where $\text{PDF}_i(z)$ is the PDF of the object $i$,
\begin{align}
  \label{eq:HPDCI}
  c_i = \sum_{z~\in~\text{PDF}_i(z)~\geqslant~\text{PDF}_i(z_\text{spec}, i)} \text{PDF}_i(z),
\end{align}
where $c_i$ is the confidence interval for the PDF of object $i$ which, ideally, should be equal to the empirical cumulative distribution function $F(c)$,
\begin{align}
  \label{eq:CRPS}
  \text{CRPS}_i = \int_{-\infty}^{+\infty} \left[ \text{CDF}_i(z) - \text{CDF}_\text{spec}(z) \right]^2 \text{d}z,
\end{align}
where $\text{CDF}_i(z)$ is the cumulative distribution function for the object $i$ and $\text{CDF}_\text{spec}(z)$ is the same for the spectroscopic redshift, represented here by a Heaviside function at $z_\text{spec}$,
\begin{align}
  \label{eq:Odds}
  \text{odds} = \int_{z_\text{peak}-\Delta z}^{z_\text{peak}+\Delta z} p(z)~\text{d}z.
\end{align}
where $z_\text{peak}$ is the photometric redshift that corresponds to the peak of the PDF and $\Delta z=0.002$.

For the single-point estimates we use the normalized mean absolute deviation \citep[$\sigma_\text{NMAD}$, ][]{Brammer+08}, the absolute bias ($\mu$), and the outlier fraction \citep[$\eta$, ][]{Ilbert2006}, defined by Eq.~\ref{eq:delta} to \ref{eq:outfrac}.
\begin{align}
  \delta z = z_\text{phot} - z_\text{spec},
  \label{eq:delta}
\end{align}

\begin{align}
  \sigma_\text{NMAD} = 1.48 \times \text{median} \left( \frac{{\delta z} - \text{median}(\delta z)}{1+z_\text{spec}} \right),
  \label{eq:nmad}
\end{align}

\begin{align}
  \mu = \text{median}(\delta z),\text{ and}
  \label{eq:bias}
\end{align}

\begin{align}
  \eta = \frac{|\delta z|}{1+z_\text{spec}} > 0.15,\text{ respectively}.
  \label{eq:outfrac}
\end{align}%

\begin{figure*}[htbp]
    \centering
    \includegraphics{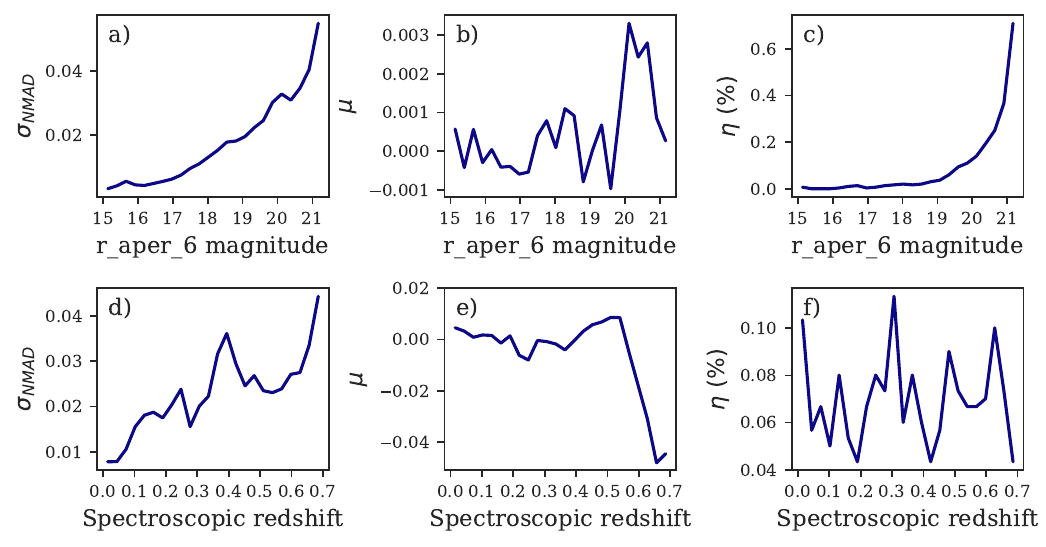}
    \caption{Results for the single-point estimates. From left to right: the scatter ($\sigma_\text{NMAD}$), absolute bias ($\mu$), and outlier fraction ($\eta$). The top row shows the results as a function of the \texttt{r\_aper\_6} magnitude and the bottom row shows the results as a function of the spectroscopic redshift.}
    \label{fig:spe_metrics}
\end{figure*}

\begin{figure*}[htbp]
    \centering
    \includegraphics{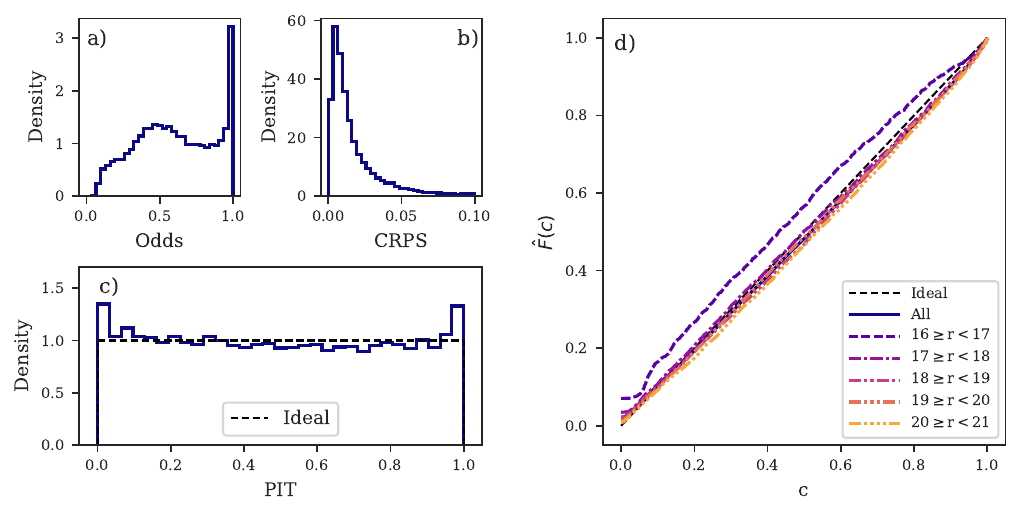}
    \caption{Results for the probability distribution functions. The panels (a), (b), and (c) show the odds, CRPS, and PIT distributions, respectively. Panel (d) shows the HPDCI curves for different bins of magnitude.}
    \label{fig:pdf_metrics}
\end{figure*}

We present the metrics calculated in a randomly selected testing sample in Fig.~\ref{fig:spe_metrics} for the single-point estimates and in Fig.~\ref{fig:pdf_metrics} for the probability distribution functions. In panels (a) and (d) of Fig.~\ref{fig:spe_metrics} we can see that the scatter of the SPEs increases as the sources get fainter. This is expected since the photometric errors increase with the magnitude, and this hinders the model's learning process. From panels (b) and (e) we note that the absolute bias is low for all ranges of magnitude and spectroscopic redshifts, except for objects with redshifts over $0.6$, where we lack enough training samples. Panels (c) and (f) illustrate that the outlier fractions remain low for all magnitude and redshift ranges.

Fig.~\ref{fig:pdf_metrics} shows the results for the PDF metrics calculated in the testing set. From panel (a) we note that most of the objects have odds above 0.5, which means that at least half of the area of their PDFs falls inside the region defined in Eq.~\ref{eq:Odds}. The CRPS can be seen in panel (b), in which we get a distribution that peaks with values close to zero, indicating a low difference between the cumulative distribution function and a real CDF, described by the Heaviside function in Eq.~\ref{eq:CRPS}. Panel (c) reflects how well-calibrated the PDFs are in terms of width, as our results follow the ideal uniform distribution. Panel (d) shows the HPDCI curves obtained for different magnitude bins in the $r$-band, and we note that, although objects with magnitudes between 16 and 17 have their PDFs with slightly lower accuracy, the curves for the other bins indicate that the PDFs can reliably represent the uncertainty in the estimates.

% \FloatBarrier
\section{Column descriptions}
\label{apd:columns}
%----------------------------------------------------------------

In this section, we describe the columns of the DR4 schemes. To simplify the notation, \{aperture\} corresponds to each of the apertures available in the \single and \dual mode photometry (\PStotal, \texttt{aper\_3}, \texttt{aper\_6}, \auto, \petro, \iso); and \{filter\} corresponds to each of the S-PLUS filters ($u$, $J0378$, $J0395$, $J0410$, $J0430$, $g$, $J0515$, $r$, $J0660$, $i$, $J0861$, $z$). We list the available DR4 schemes below:

%----------------------------------------------------------------
%\subsection{DR4\_dual}
%----------------------------------------------------------------

\begin{itemize}
    \item The dual mode photometry schema (\texttt{DR4\_dual}) contains 13 tables: one for the columns related to the catalogue produced from the detection image, and one for the catalogue of each of the 12 filters. Tables~\ref{tab:detection_image} and \ref{tab:dual_filter} list the characteristics for each of the columns in the \dual-mode schema.

    \item The single-mode aperture photometry schema (\texttt{DR4\_single}) contains 12 tables, corresponding to each S-PLUS filter. Table~\ref{tab:single_filter} lists all the parameters available.

    \item The PSF photometry schema (\texttt{DR4\_psf}) also contains 12 tables, corresponding to each S-PLUS filter. Table~\ref{tab:psf_filter} lists the characteristics of each column for this mode.

    \item Table~\ref{tab:qso_class} lists the descriptions of the columns present in the star/galaxy/quasar classification VAC.

    \item Tables~\ref{tab:photoz_columns} and \ref{tab:qso_z} describe the columns of the galaxies and quasars photometric redshifts VAC, respectively.

    \item Tables~\ref{tab:VAC_CNN} and \ref{tab:VAC_ANN} describe the columns of the Convolutional neural network and the \texttt{SPHINX} stellar parameters VACs, respectively. For comparison, \citet{2021ApJ...912..147W} computed 0.25 and 0.35 dex residuals for [Fe/H] and A(C), which can be considered by the user when querying the DR4 data.

\end{itemize}

\begin{table*}[!]
\caption{Description of the columns of the detection image catalogue for the \dual mode photometry.}
\label{tab:detection_image}
\centering
\begin{tabular}{@{}lp{10cm}l@{}}
\toprule
Column & Description & Unit \\ \midrule
Field & Name of S-PLUS field & \\
ID & Object ID in a given field & \\
ID\_RA & Right ascension (J2000) associated with this ID & [deg]\\
ID\_DEC & Declination (J2000) associated with this ID & [deg] \\
PHOT\_ID\_dual & Same as DET\_ID\_dual & \\
PHOT\_ID\_RA\_dual & Same as RA & [deg] \\
PHOT\_ID\_DEC\_dual & Same as DEC & [deg] \\
DET\_ID\_dual & Generated ID for the source in the detection catalogue &  \\
RA & Right ascension of barycenter (J2000) of the source in this filter & [deg] \\
DEC & Declination of barycenter (J2000) of the source in this filter & [deg] \\
SEX\_NUMBER\_DET & SExtractor's Running object number & \\
SEX\_FLAGS\_DET & SExtractor's Extraction flags & \\
calib\_strat & Name given to calibration strategy used in this field &      \\
X & Object position along x & [pixel] \\
Y & Object position along y & [pixel] \\
A & Profile RMS along major axis & [deg] \\
B & Profile RMS along minor axis & [deg] \\
THETA & Position angle (CCW/World-x) & \\
ELONGATION & A\_IMAGE/B\_IMAGE & \\
ELLIPTICITY & 1 - B\_IMAGE/A\_IMAGE &      \\
ISOarea & Isophotal area above ANALYSIS\_THRESH & [deg$^2$] \\
KRON\_RADIUS & 	Kron radius in units of A or B & \\
PETRO\_RADIUS & Petrosian apertures in units of A or B & \\
FLUX\_RADIUS\_20 & Radius enclosing 20\% of the total flux & \\
FLUX\_RADIUS\_50 & Radius enclosing 50\% of the total flux & \\
FLUX\_RADIUS\_70 & Radius enclosing 70\% of the total flux & \\
FLUX\_RADIUS\_90 & Radius enclosing 90\% of the total flux & \\
CLASS\_STAR & S/G classifier output (1: star; 0: non-star) & \\
FWHM & FWHM assuming a Gaussian core & [deg] \\
FWHM\_n & Normalized FWHM assuming a Gaussian core & \\
MU\_MAX\_INST & Instrumental Peak surface brightness above the background & [instrumental] \\
MU\_THRESHOLD\_INST & Instrumental Detection threshold above background & [instrumental] \\
BACKGROUND & Instrumental background at centroid position & [instrumental] \\
THRESHOLD & Instrumental detection threshold above background & [instrumental] \\
s2n\_DET\_\{aperture\} & signal to noise ratio of {aperture} measurement & \\
EBV\_SCH & extinction E\_\{B-V\} given by \citet{Schlegel+98} maps for the source's RA DEC' & [mag]\\
\bottomrule
\end{tabular}
\end{table*}

\begin{table*}[!]
\caption{Description of the columns of the catalogues for the \dual-mode photometry, where \{filter\} represents any of the 12 S-PLUS bands ($u$, $J0378$, ...), and \{aperture\} represent any of the apertures (\texttt{auto}, \texttt{petro}, \texttt{aper\_3}).}
\label{tab:dual_filter}
\centering
\begin{tabular}{@{}lll@{}}
\toprule
Column & Description & Unit \\ \midrule
Field & Name of S-PLUS field & \\
ID & Object ID in a given field & \\
ID\_RA & Right ascension (J2000) associated with this ID & [deg]\\
ID\_DEC & Declination (J2000) associated with this ID & [deg] \\
PHOT\_ID\_dual & Same as DET\_ID\_dual & \\
PHOT\_ID\_RA\_dual & Same as RA & [deg] \\
PHOT\_ID\_DEC\_dual & Same as DEC & [deg] \\
\{filter\}\_ID\_dual & In dual mode, same as DET\_ID\_dual & \\
RA & Right ascension of barycenter (J2000) of the source in this filter & [deg] \\
DEC & Declination of barycenter (J2000) of the source in this filter & [deg] \\
SEX\_NUMBER\_\{filter\} & SExtractor's Running object number & \\
SEX\_FLAGS\_\{filter\} & SExtractor's Extraction flags & \\
calib\_strat & Name given to calibration strategy used in this field &      \\
FWHM\_\{filter\} & FWHM assuming a Gaussian core in this filter & [deg] \\
FWHM\_n\_\{filter\} & Normalized FWHM assuming a Gaussian core & \\
MU\_MAX\_\{filter\} & Peak surface brightness above background in this filter & [mag/arcsec$^2$] \\
MU\_THRESHOLD\_\{filter\} & Detection threshold above background in this filter & [mag/arcsec$^2$] \\
BACKGROUND\_\{filter\} & Instrumental background at centroid position in this filter & [instrumental] \\
THRESHOLD\_\{filter\} & Instrumental detection threshold above background in this filter & [instrumental] \\
\{filter\}\_\{aperture\} & AB-Calibrated magnitude for the \{aperture\} measurement & [AB mag] \\
e\_\{filter\}\_\{aperture\} & Error for \{filter\}\_\{aperture\} magnitude & [AB mag] \\
s2n\_\{aperture\}\_\{filter\} & Signal to noise ratio of \{aperture\} measurement &\\
\bottomrule
\end{tabular}
\end{table*}

%----------------------------------------------------------------
%\subsection{DR4\_single}
%----------------------------------------------------------------

\begin{table*}[!]
\caption{Description of the columns of the catalogues for the \single-mode photometry, where \{filterz\} represents any of the 12 S-PLUS bands ($u$, $J0378$, ...), and \{aperture\} represent any of the apertures (\texttt{auto}, \texttt{petro}, \texttt{aper\_3}).}
\label{tab:single_filter}
\centering
\begin{tabular}{@{}lp{10cm}l@{}}
\toprule
Column & Description & Unit \\ \midrule
Field & Name of S-PLUS field & \\
ID & Object ID in a given field & \\
ID\_RA & Right ascension (J2000) associated with this ID & [deg]\\
ID\_DEC & Declination (J2000) associated with this ID & [deg] \\
PHOT\_ID\_single & Generated single mode photometry ID & \\
PHOT\_ID\_RA\_single & Right ascension (J2000) associated with this PHOT\_ID & [deg] \\
PHOT\_ID\_DEC\_single & Declination (J2000) associated with this PHOT\_ID & [deg] \\
\{filter\}\_ID\_single & Generated ID for the detection in this filter in single mode & \\
RA\_\{filter\} & Right ascension of barycenter (J2000) of the source in this filter & [deg] \\
DEC\_\{filter\} & Declination of barycenter (J2000) of the source in this filter & [deg] \\
SEX\_NUMBER\_\{filter\} & SExtractor's Running object number & \\
SEX\_FLAGS\_\{filter\} & SExtractor's Extraction flags & \\
calib\_strat & Name given to calibration strategy used in this field & \\
X\_\{filter\} & Object position along x in this filter & [pixel] \\
Y\_\{filter\} & Object position along y in this filter & [pixel] \\
A\_\{filter\} & Profile RMS along major axis in this filter & [deg] \\
B\_\{filter\} & Profile RMS along minor axis in this filter & [deg] \\
THETA\_\{filter\} & Position angle (CCW/World-x) in this filter & [deg] \\
ELONGATION\_\{filter\} & A\_IMAGE/B\_IMAGE & \\
ELLIPTICITY\_\{filter\} & 1 - B\_IMAGE/A\_IMAGE in this filter &      \\
ISOarea\_\{filter\} & Isophotal area above ANALYSIS\_THRESH in this filter & [deg$^2$] \\
KRON\_RADIUS\_\{filter\} & 	Kron radius in units of A or B in this filter & \\
PETRO\_RADIUS\_\{filter\} & Petrosian apertures in units of A or B in this filter & \\
FLUX\_RADIUS\_20\_\{filter\} & Radius enclosing 20\% of the total flux in this filter & \\
FLUX\_RADIUS\_50\_\{filter\} & Radius enclosing 50\% of the total flux in this filter & \\
FLUX\_RADIUS\_70\_\{filter\} & Radius enclosing 70\% of the total flux in this filter & \\
FLUX\_RADIUS\_90\_\{filter\} & Radius enclosing 90\% of the total flux in this filter & \\
CLASS\_STAR\_\{filter\} & S/G classifier output (1: star; 0: non-star) in this filter & \\
FWHM\_\{filter\} & FWHM assuming a Gaussian core in this filter & [deg] \\
FWHM\_n\_\{filter\} & Normalized FWHM assuming a Gaussian core & \\
MU\_MAX\_\{filter\} & Peak surface brightness above background in this filter & [mag/arcsec$^2$] \\
MU\_THRESHOLD\_\{filter\} & Detection threshold above background in this filter & [mag/arcsec$^2$] \\
BACKGROUND\_\{filter\} & Instrumental background at centroid position in this filter & [instrumental] \\
THRESHOLD\_\{filter\} & Instrumental detection threshold above background in this filter & [instrumental] \\
\{filter\}\_\{aperture\} & AB-Calibrated magnitude for the \{aperture\} measurement & [AB mag] \\
e\_\{filter\}\_\{aperture\} & Error for \{filter\}\_\{aperture\} magnitude & [AB mag] \\
s2n\_\{aperture\}\_\{filter\} & Signal to noise ratio of \{aperture\} measurement & \\
EBV\_SCH & Extinction E\_\{B-V\} given by \citet{Schlegel+98} maps for the source's RA\_\{filter\}, DEC\_\{filter\} & mag \\
\bottomrule
\end{tabular}
%\end{adjustbox}
\end{table*}

%----------------------------------------------------------------
%\subsection{dr4\_psf}
%----------------------------------------------------------------

\begin{table*}[!]
\caption{Description of the columns of the catalogues for the \PSF-mode photometry, where \{filter\} represents any of the 12 S-PLUS bands ($u$, $J0378$, ...), and \{aperture\} represent any of the apertures (\texttt{auto}, \texttt{petro}, \texttt{aper\_3}).}
\label{tab:psf_filter}
\centering
\begin{tabular}{@{}lp{10cm}l@{}}
\toprule
Column & Description & Unit \\ \midrule
Field & Name of S-PLUS field & \\
ID & Object ID in a given field & \\
ID\_RA & Right ascension (J2000) associated with this ID & [deg]\\
ID\_DEC & Declination (J2000) associated with this ID & [deg] \\
PHOT\_ID\_psf & Generated single mode photometry ID & \\
PHOT\_ID\_RA\_psf & Right ascension (J2000) associated with this PHOT\_ID & [deg] \\
PHOT\_ID\_DEC\_psf & Declination (J2000) associated with this PHOT\_ID & [deg] \\
RA\_\{filter\} & Right ascension of barycenter (J2000) of the source in this filter & [deg] \\
DEC\_\{filter\} & Declination of barycenter (J2000) of the source in this filter & [deg] \\
DoPHOT\_Star\_number\_\{filter\} & DoPHOT star number for this filter & \\
calib\_strat & Name given to calibration strategy used in this field & \\
X\_\{filter\} & Object position along x in this filter & [pixel] \\
Y\_\{filter\} & Object position along y in this filter & [pixel] \\
CLASS\_STAR\_\{filter\} & S/G classifier output (1: star; 0: non-star) in this filter & \\

\{filter\}\_psf & AB-Calibrated PSF-photometry magnitude & [AB mag] \\
e\_\{filter\}\_psf & Error for \{filter\}\_PSF magnitude & [AB mag] \\
s2n\_psf\_\{filter\} & Signal to noise ratio of PSF-photometry measurement & \\
EBV\_SCH & Extinction E\_\{B-V\} given by \citet{Schlegel+98} maps for the source's RA\_\{filter\}, DEC\_\{filter\} & mag \\
\bottomrule
\end{tabular}
\end{table*}

%----------------------------------------------------------------
%\section{VAC column descriptions}
%\label{apd:VAC_col_des}

\begin{table*}[!]
\centering
\caption{Columns in the DR4 star-galaxy-quasar VAC.}
\label{tab:qso_class}
\begin{tabular}{@{}p{4cm}p{12cm}@{}}
\toprule
Column & Description \\
\midrule
\texttt{CLASS} & 0 = QSO, 1 = STAR, 2 = GALAXY\\
\texttt{PROB\_STAR} & Probability of a source being a star \\
\texttt{PROB\_QSO} & Probability of a source being a quasar \\
\texttt{PROB\_GAL} & Probability of a source being a galaxy \\
\texttt{model\_flag} & 0 = classified with S-PLUS + AllWISE; \newline 1 = classified with S-PLUS only \\
\bottomrule
\end{tabular}
\end{table*}
%----------------------------------------------------------------
%\subsection{Galaxy photometric redshifts}
%----------------------------------------------------------------

\begin{table*}[!]
\caption{Description of the columns in the VAC photometric redshifts for galaxies.}
\label{tab:photoz_columns}
\centering
    \begin{tabular}{@{}p{4cm}p{12cm}@{}}
        \toprule
        Column     & Description                                                                                                   \\ \midrule
        ID              & Identifier of the object in the dual photometry catalogue \\ 
        RA              & Right ascension [degrees]                                                                                     \\ 
        DEC             & Declination [degrees]                                                                                         \\ 
        zml             & Single-point estimate of the photometric redshift, calculated as a maximum-a-posteriori estimate from the PDF \\ 
        zml\_2.5q        & 2.5\% percentile of the CDF of the object                                                                      \\ 
        zml\_16q         & 16\% percentile of the CDF of the object                                                                       \\ 
        zml\_84q         & 84\% percentile of the CDF of the object                                                                       \\ 
        zml\_97.5q       & 97.5\% percentile of the CDF of the object                                                                     \\ 
        odds            & Odds of the object, calculated as in Benitez et al. 2000                                                      \\ 
        pdf\_err         & Width of the PDF                                                                                              \\ 
        pdf\_peaks       & Number of peaks detected in the PDF                                                                           \\ 
        zml\_second\_peak & Single-point estimate of the redshift of the second peak, if one is detected                                  \\ 
        pdf\_weights     & Seven components of weights for the PDF mixture                                                               \\ 
        pdf\_means       & Seven components of means for the PDF mixture                                                                 \\ 
        pdf\_stds        & Seven components of standard deviation for the PDF mixture                                                    \\
        \bottomrule
    \end{tabular}
\end{table*}

%----------------------------------------------------------------
%\subsection{Quasar photometric redshifts}
%----------------------------------------------------------------

\begin{table*}
\centering
\caption{Description of the columns provided in DR4 regarding quasar photometric redshifts, along with S-PLUS ID, RA, and Dec. All columns are type \texttt{float64}. }
\label{tab:qso_z}
\begin{tabular}{@{}p{4cm}p{12cm}@{}}
\toprule
Column & Description \\ \midrule
z\_rf      &  Photo-z estimated with \rf{}         \\
z\_bmdn\_peak      &  Photo-z estimated with \bnn{} (peak of the PDF)           \\ 
z\_flex\_peak      &   Photo-z estimated with \flex{} (peak of the PDF)          \\
z\_mean        &    Average of \{z\_rf, z\_bmdn\_peak, and z\_flex\_peak\}         \\ 
z\_std      &   Standard deviation of \{z\_rf, z\_bmdn\_peak, and z\_flex\_peak\}         \\
n\_peaks\_bmdn        &   Number of peaks for \bnn{}'s PDF          \\ 
z\_bmdn\_pdf\_weight\_[0-6]     & Weight of the [0-6]-th Gaussian distribution estimated with \bnn{}          \\
z\_bmdn\_pdf\_mean\_[0-6]       &  Mean of the [0-6]-th Gaussian distribution estimated with \bnn{}          \\ 
z\_bmdn\_pdf\_std\_[0-6]       &  Standard deviation of the [0-6]-th Gaussian distribution estimated with \bnn{}          \\ 
z\_flex\_pdf\_[1-200]       &  Probability for the [1-200]-th redshift within the interval [0.058, 6.999] estimated with \flex{}          \\ 
\bottomrule
\end{tabular}
\end{table*}

%----------------------------------------------------------------
%\subsection{Stellar parameters using CNN}
%----------------------------------------------------------------

\begin{table*}[!]
\centering
\caption{Columns in the DR4 CNN stellar parameters VAC. }
\label{tab:VAC_CNN}
\begin{tabular}{@{}p{4cm}p{12cm}@{}}
\toprule
Column & Description \\
\midrule
ID & Identifier in S-PLUS DR4 \\ 
RA & Right ascension [degrees] \\ 
DEC & Declination [degrees] \\
GAIADR3\_SOURCE\_ID & SOURCE ID from Gaia DR3  \\
\texttt{TEFF\_CNN} & Effective temperature [K] \\
\texttt{FEH\_CNN} & Metallicity \\
\texttt{LOGG\_CNN} & Surface Gravity \\
\bottomrule
\end{tabular}
\end{table*}

%----------------------------------------------------------------
%\subsection{Stellar parameters using SPHINX}
%----------------------------------------------------------------

\begin{table*}[!]
\centering
\caption{Columns in the DR4 \texttt{SPHINX} stellar parameters VAC. Each reported parameter corresponds to the average of the ANN sub-unit estimates, weighted by the validation scores. The reported errors correspond to the error propagation between the scatter among the individual ANN sub-units and the standard deviation of the spectroscopic residuals of the validation stars. For the effective temperature, we computed a residual of 120 K for the SEGUE DR12 adopted values. The cases when \texttt{SPHINX} does not provide an estimate for the parameter are represented as $-9999$.}
\label{tab:VAC_ANN}
\begin{tabular}{@{}p{4cm}p{12cm}@{}}
\toprule
Column & Description \\
\midrule
ID & Identifier of the object in the dual photometry catalogue \\ 
RA & Right ascension [degrees] \\ 
DEC & Declination [degrees] \\
\texttt{NET\_FEH} & Metallicity  \\
\texttt{NET\_FEH\_ERR} & Metallicity uncertainty \\
\texttt{NET\_ARRAY\_FEH\_FLAG} & Number of ANN sub-units used in the metallicity estimation \\
\texttt{NET\_AC} & Absolute carbon abundance \\
\texttt{NET\_AC\_ERR} & Absolute carbon abundance uncertainty  \\
\texttt{NET\_ARRAY\_AC\_FLAG} & Number of ANN sub-units used in the absolute carbon abundance estimation \\
\texttt{NET\_TEFF} & Effective temperature [K] \\
\texttt{NET\_TEFF\_ERR} & Effective temperature uncertainty [K]  \\
\texttt{NET\_ARRAY\_TEFF\_FLAG} & Number of ANN sub-units used in the effective temperature estimation \\
\texttt{NET\_OVERLAP\_FLAG} & 0 = based on  the architecture of only one temperature regime; \newline 1 = [Fe/H] and A(C) values are the averages between the measurements from the two temperature regimes \\
\bottomrule
\end{tabular}
\end{table*}

%\begin{thebibliography}{99}
%\end{thebibliography}

%%%%%%%%%%%%%%%%%%%%%%%%%%%%%%%%%%%%%%%%%%%%%%%%%%

% Don't change these lines
% \bsp	% typesetting comment
% \label{lastpage}
\end{document}